\documentclass[prd,amsmath,amssymb,longbibliography,superscriptaddress,twocolumn,nofootinbib,10pt]{revtex4}

\usepackage{dcolumn}
\usepackage{bm}
\usepackage{amssymb}
\usepackage[T1]{fontenc}
\usepackage[utf8]{inputenc}
\usepackage{booktabs}
\usepackage{amsmath}
\usepackage{multirow}
\usepackage{graphicx} 
\RequirePackage{xcolor}

\usepackage[colorlinks,linkcolor=red,anchorcolor=red,citecolor=blue,hyperfootnotes=true]{hyperref}
\newcommand{\beq}[1]{\begin{equation}\label{#1}}
 \newcommand{\eeq}{\end{equation}}
 \newcommand{\bea}{\begin{eqnarray}}
 \newcommand{\eea}{\end{eqnarray}}

\def\({\left(}
\def\){\right)}
\def\[{\left[}
\def\]{\right]}

\begin{document}

\title{Identifying Microlensing by Compact Dark Matter through Diffraction Patterns in Gravitational Waves with Machine Learning}

\author{Ao Liu}
\affiliation{Department of Physics, and Collaborative Innovation Center for Quantum Effects and Applications, Hunan Normal University, Changsha 410081, China;}
\affiliation{College of Information Science and Engineering, Hunan Normal University, Changsha, Hunan 410081, People's Republic of China;}

\author{Tonghua Liu}
\email{liutongh@yangtzeu.edu.cn;}
\affiliation{School of Physics and Optoelectronic Engineering, Yangtze University, Jingzhou, 434023, China;}

\author{Dejiang Li}
\affiliation{Department of Physics, and Collaborative Innovation Center for Quantum Effects and Applications, Hunan Normal University, Changsha 410081, China;}
\affiliation{College of Information Science and Engineering, Hunan Normal University, Changsha, Hunan 410081, People's Republic of China;}

\author{Cuihong Wen}
\email{cuihongwen@hunnu.edu.cn;}
\affiliation{Department of Physics, and Collaborative Innovation Center for Quantum Effects and Applications, Hunan Normal University, Changsha 410081, China;}
\affiliation{College of Information Science and Engineering, Hunan Normal University, Changsha, Hunan 410081, People's Republic of China;}

\author{Jieci Wang}
\email{jcwang@hunnu.edu.cn;}
\affiliation{Department of Physics, and Collaborative Innovation Center for Quantum Effects and Applications, Hunan Normal University, Changsha 410081, China;}

\author{Kai Liao}
\email{liaokai@whu.edu.cn;}
\affiliation{School of Physics and Technology, Wuhan University, Wuhan 430072, China;}

\author{Jiaxing Cui}
\affiliation{Department of Physics, and Collaborative Innovation Center for Quantum Effects and Applications, Hunan Normal University, Changsha 410081, China;}
\affiliation{College of Information Science and Engineering, Hunan Normal University, Changsha, Hunan 410081, People's Republic of China;}

\author{Huan Zhou}
\affiliation{School of Physics and Optoelectronic Engineering, Yangtze University, Jingzhou, 434023, China;}

\begin{abstract}
Gravitational wave lensing, particularly microlensing by compact dark matter (DM), offers a unique avenue to probe the nature of dark matter. However, conventional detection methods are often computationally expensive, inefficient, and sensitive to waveform systematics. In this work, we introduce the Wavelet Convolution Detector (WCD), a deep learning framework specifically designed to identify wave-optics diffraction patterns imprinted in gravitationally lensed signals.  The WCD integrates multi-scale wavelet analysis within residual convolutional blocks to efficiently extract time-frequency interference structures, and is trained on a realistically generated dataset incorporating compact DM mass functions and astrophysical lensing probabilities. This work is the first machine learning-based approach capable of identifying such wave-optics signatures in lensed gravitational waves. Tested on  simulated binary black hole events, the model achieves 92.2\% accuracy (AUC=0.965), with performance rising to AUC$\sim$0.99 at high SNR. Crucially, it maintains high discriminative power across a wide range of lens masses without retraining, demonstrating particular strength in the low-impact-parameter and high-lens-mass regimes where wave-optics effects are most pronounced. Compared to Bayesian inference, the WCD provides orders-of-magnitude faster inference, making it a scalable and efficient tool for discovering  compact DM through lensed gravitational waves in the era of third-generation detectors.


\end{abstract}

\maketitle

\section{Introduction}           
Since the first detection of GW150914 in 2015~\cite{PhysRevLett.116.061102, k38}, gravitational-wave astronomy has rapidly developed into a fundamental discipline for probing extreme astrophysical environments~\cite{k2, k39}. With ongoing enhancements in the sensitivity of ground-based detectors such as LIGO, Virgo, and KAGRA~\cite{Acernese_2015, 10.1093/ptep/ptaa125}, along with improved frequency coverage and source localization accuracy, it has become possible to observe the coalescence of binary black holes, neutron star mergers, and potential signatures of new physics~\cite{Abbott_2019, PhysRevLett.119.161101, k4}. This window of observation not only validates general relativity in the strong-field regime~\cite{Barack_2019, Berti_2015}, but also offers an unprecedented experimental foundation for measuring cosmological parameters, investigating stellar astrophysics, and constraining the equation of state of dense nuclear matter~\cite{Abbott_2017, ZhuWei2021ESaT, Annala_2020}.

In recent years, the field has increasingly turned its attention to the more complex phenomenon of gravitational wave lensing~\cite{k6,Oguri_2019,Liao:2022gde}. This process, which previously enabled the first detections of exoplanets through microlensing, also offers some of the most compelling evidence for the existence of dark matter substructures~\cite{Vegetti_2009, Hezaveh_2017}. When gravitational waves propagate near massive objects—such as stars, intermediate-mass black holes, or dark matter subhalos—the resulting spacetime curvature bends the wavefront. In scenarios where the GW wavelength is comparable to the Schwarzschild radius of the lens, wave diffraction becomes significant~\cite{Nakamura_2009, Takahashi_2003}, leading to frequency-dependent phase shifts, amplitude amplification in specific bands, and distinct signal splitting in the time-frequency domain~\cite{Dai_2018, PhysRevLett.122.062701}. These diffractive imprints open a unique window for detecting optically invisible compact objects, constraining the properties of dark matter substructure, and testing gravitational theories beyond general relativity. For instance, measurements of lensing time delays can be used to independently constrain the Hubble constant~\cite{2017NatCo...8.1148L,k8}, while the statistics of strongly lensed events provide insights into the demographics of lensing galaxies and the clustering of dark matter~\cite{k9}.

However, current identification frameworks face three major bottlenecks as data volumes increase: matched filtering is highly sensitive to waveform template mismatch, resulting in significant sensitivity loss when templates are imperfect~\cite{PhysRevD.60.022002, Cornish_2015}; Bayesian parameter estimation, while providing complete posterior distributions, demands tens of thousands of CPU hours per event and becomes computationally intractable at scale~\cite{PhysRevD.91.042003}; and most critically, conventional techniques exhibit markedly reduced performance under low signal-to-noise ratios (SNR) or complex lensing geometries~\cite{Janquart2022AnalysesOO}. These constraints become especially acute in light of upcoming observational runs. In the third-generation detector era, event numbers are projected to reach $\mathcal{O}(10^5\text{--}10^6)$, with a lensing rate around 0.3\%, necessitating $\mathcal{O}(10^{10}\text{--}10^{12})$ pair-wise comparisons~\citep{k40,k41}. The prohibitive computational cost of traditional Bayesian inference renders it infeasible for real-time lensed candidate identification, underscoring the urgent need for more efficient, scalable, and robust detection strategies.

The recent advancements in deep learning have demonstrated significant potential in the field of gravitational wave lensing identification~\cite{Oguri_2019}. Neural networks, with their powerful capability for high-dimensional feature extraction and noise suppression~\cite{PhysRevLett.120.141103, Green_2021, PhysRevLett.124.041102}, can rapidly and accurately distinguish between lensed and unlensed signals~\cite{Gao2023, Collett2015}. Their inference speed far exceeds that of traditional Bayesian methods, and they exhibit excellent generalization across multi-parameter spaces. Several studies have endeavored to apply deep learning to this problem. \citet{Magare:2024wje} significantly improved classification accuracy by analyzing time--frequency representations of gravitational wave signals, such as Q-transforms and sine-Gaussian representations (SGP). \citet{Goyal2021} combined XGBoost with a DenseNet201 model to extract discriminative features from time--frequency maps for classification. \citet{Kim:2020xkm} also employed deep learning methods for effective identification of lensed gravitational waves. In terms of theoretical modeling, \citet{Lin2023} emphasized the important role of wave-optical effects in the lensing process of gravitational waves, while \citet{Pagano2020} developed the \textsf{lensingGW} Python package for studying macromodels containing multiple lensing bodies. In addition, some researchers have attempted to apply unsupervised learning \cite{Sheng2022} and other machine learning methods \cite{KeerthiVasan2023,2024MNRAS.532.4842C} to the problem of lens identification.

Although existing studies have made progress in lensed/unlensed classification, most approaches remain confined to the geometric-optics approximation, lack targeted modeling of wave-optical effects (particularly diffraction), and exhibit insufficient cross-detector generalization capability~\cite{10.1093/mnras/stab579}. Therefore, developing identification methods that incorporate wave-optical principles and possess strong cross-platform transferability remains an important direction for future research. This paper proposes a Diffraction-Aware Gravitational Wavelet Convolutional Detector (WCD). Its core innovation lies in explicitly modeling diffraction features in the time-frequency domain using a multi-scale wavelet encoder~\cite{finder2024wavelet}, combined with depthwise residual convolutions to focus on critical diffraction regions~\cite{he2016deep}, significantly reducing model parameters. Furthermore, leveraging a synthetic training set generated based on compact dark matter of mass distributions and lens redshift distributions, we introduce a domain adaptation strategy to ensure model robustness on unseen real data.

The outline of this paper is organized as follows. Section~2  reviews the physical foundations of GW lensing. Section 3 details simulated data generation and experimental setup. We presents the WCD network architecture and training strategy in Section 4. The final  experimental results and ablation studies are reported in  Section 5. Section 6 concludes the paper and outlines future work. We assume the a flat $\Lambda$CDM cosmology and the Hubble parameter $H_0 = 70.0~\rm km/s/Mpc$ and present density parameter of dark matter $\Omega_{\rm DM}=0.3$, and use the units of $c = G = 1$.

\section{Simulation for Microlensing-Induced Diffraction Patterns in Gravitational Wave Signals}
\subsection{Gravitational Wave  in wave Optics lensed by the Point Mass Lenses}
\begin{figure*}
  \centering
  \includegraphics[width=0.9\linewidth]{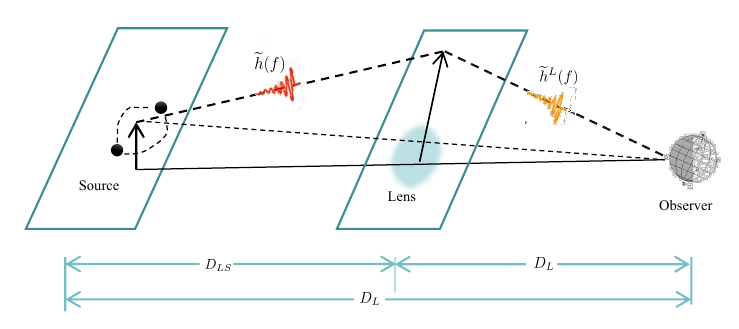}
  \caption{Gravitational waveforms illustrating the lensing effect of point mass lenses on gravitational waves. The figure shows the original gravitational wave signal $\widetilde{h}(f)$ and the lensed gravitational wave signal $\widetilde{h}^{\rm L}(f)$, incorporating key distances such as the luminosity distance $D_{\rm L}$ and the distance from the lens to the source $D_{\rm LS}$. These elements help visualize how gravitational waves are affected as they pass through the spacetime curvature caused by a point mass lens.}
  \label{fig:GW_lens}
\end{figure*}

The main effects of gravitational lensing on gravitational wave signals arise from amplitude amplification, time delay of the signal reaching the Earth (for multiple images), and phase modulations~\cite{k14,k43}.
It is easy for us to know the formula for calculating the magnification factor $F(f)$ based on the relationship between lenses and unlenses cases
\begin{equation}
\widetilde{h}^{\rm L}(f) = F(f)\widetilde{h}(f),
\end{equation}
where $\widetilde{h}^{\rm L}(f)$ and $\widetilde{h}(f)$ are the lensed and unlensed gravitational wave amplitudes, respectively. Figure \ref{fig:GW_lens}  presents a schematic diagram illustrating the principle of simple lensing gravitational waves.
Depending on the different wavelengths of gravitational waves $\lambda_{\rm GW}$, the optical effects of lensing can be categorized into two regimes: when  $\lambda_{\rm GW}$ is much smaller than the Schwarzschild radius of the lens, the geometric optics limit is valid. On the other hand, when  $\lambda_{\rm GW}$ is larger than the Schwarzschild radius of the lens, diffraction effects become significant, and the wave optics limit applies in this regime. The applicable regions for each optical limit can also be determined based on the lens mass $m_{\rm len}$. For instance, \citet{Takahashi_2003} discussed how diffraction effects become important when $m_{lens}$ satisfies certain conditions, i.e.,
\begin{equation}
m_{\rm len}\lesssim10^8 M_{\odot}\bigg(\frac{f}{{\rm mHz}}\bigg)^{-1}.
\end{equation}

In this work, we  consider the third-generation gravitational wave detector, the Einstein Telescope (ET), whose sensitive frequency range is $1-10^3$ Hz. This means that when the $m_{\rm len}$ is less than $10^5 M_{\odot}$, the wave-optical theory should be used to describe the lensing gravitational waves. For lenses within this quality range, the representative celestial bodies could be stellar-mass black holes, intermediate-mass black holes, or dense celestial matter such as  compact dark matter structures and primordial black holes. Regarding the properties of these celestial bodies, there has not been sufficient research conducted on them yet. Therefore,  we assume a point-mass lensing model and consider compact dark matters (DMs)  as lenses. We will employ a realistic compact DM masses distribution model.
The specific details of the simulation will be introduced in the next section. Obviously, our research scope is wave optics, and within this frequency range and quality range, there is only diffraction effect.

Diffraction effect results are as a superposition of all possible waves on the lens plane that have different time delays corresponding to different phases.
If we assume that $D_{\rm L}$, $D_{\rm S}$, and $D_{\rm LS}$ represent the angular diameter distances from the observer to the lens, observer to the light source, and lens to light source, respectively. Gravitational waves pass through a thin lens plane with a surface mass density $\Sigma(\xi)$, and the amplification factor $F(f)$ of the observer's position can be derived based on diffraction integration in wave optics~\cite{Sun_2023}
\begin{equation}
F(f) = \frac{D_{\rm S} \xi_0^2(1+z_{\rm l})}{D_{\rm L} D_{\rm LS}} \frac{f}{i}  \int d^2x \, \exp\left[ 2\pi i f t_{\rm d}(x, y) \right],\label{eq3}
\end{equation}
where $x = \xi / \xi_0$ is the position vector of the dimensionless lens plane, $\xi_0$ is the arbitrary normalization constant of the length, and $y = D_{\rm L}\eta/ D_{\rm S}\xi_0$ represents the dimensionless source position vector ($\eta $ is the true positional offset of the source in the source plane). The $t_{\rm d} (x, y)$ represents the total travel time difference from the source position $y$ through the lens point $x$ to the observer.
The complete time delay expression is:
\begin{equation}
t_{\rm d}(x, y) = \frac{D_{\rm S} \xi_0^2(1 + z_{\rm l})}{D_{\rm L} D_{\rm LS}}  \left[ \frac{1}{2}|x - y|^2 - \psi(x) + \phi_m(y) \right].
\end{equation}
For a point mass lens model, its dimensionless lensing  potential is given by $\psi(x) = \ln x$, and $\phi_{\rm m}(y)= \frac{1}{2}(x_m - y)^2 - \ln x_{\rm m}$ is chosen so that the minimum arrival time is zero. Thus, we have $x_{\rm m} = (y + \sqrt{y^2 + 4})/2$.

Under this situation,  the Eq. (\ref{eq3}) has an analytical solution:
\begin{eqnarray}\label{eq5}
F(f)&=&\exp\left( \frac{\pi w}{4} + i \frac{w}{2} \left[ \ln\left( \frac{w}{2} \right) - 2\phi_m(y) \right] \right) \\\nonumber
&\times&\Gamma\left(1 - i \frac{w}{2}\right) F_1\left( i \frac{w}{2}; 1; i \frac{w y^2}{2} \right),
\end{eqnarray}
where the dimensionless frequency is $w = {8\pi M_{\rm Lz} f}$ with $M_{\rm Lz} =m_{\rm len} (1 + z_{\rm l})$ being the redshifted lens mass, and  $F_1$ is the confluent hyper-geometric function. The typical time delay $\Delta t_{\rm d}\sim 2\times 10^3$ $(M_{\rm Lz}/10^8 M_{\odot})~{\rm s}$, which means that there is only a single image in our study.

The lensing models we have chosen are simple (analytical solutions are accessible) but sufficient to produce qualitative lensing features, In the single-image mode, the modulation amplitude caused by diffraction is relatively small~\cite{Takahashi_2003}. Usually, the amplitude modulation does not exceed a few percent in fraction, and the overall distortion of the waveform is moderate. However, we will prove that as long as the signal-to-noise ratio (SNR) of the matched filtering is sufficiently high, the artificial neural network trained by machine learning methods for the time-frequency domain graph of gravitational waves can effectively distinguish the distorted waveform due to diffraction from the unlensed waveform.

\section{simulation for lensed GW events}
This work primarily focuses on distinguishing lensed from unlensed GW events using a  machine learning-based approach, based on the diffraction effects arising from gravitational lensing. We will provide a detailed description of the entire data preparation process, from the generation of simulated GWs to the preparation of time-frequency representation samples input into the machine learning model.

\subsection{The mass distribution of lenses and generating for lensed GW events}\label{sec3-3}
\textbf{Generating BBH events:} The gravitational wave sources redshift distribution is drawn from the phenomenological fit $R_{\rm BBH}$ to the population synthesis rate
\begin{equation}\label{eq3-1-1}
P_{\rm BBH}(z_{\rm s})=\frac{1}{Z_{\rm BBH}}\frac{R_{\rm BBH}(z_{\rm s})}{1+z_{\rm s}}\frac{dV_{\rm c}}{dz_{\rm s}},
\end{equation}
where BBH represents binary black holes,  $Z_{\rm BBH}$ is the normalization constant,  $R_{\rm BBH}$ is the merger rate per unit of volume in the source frame and takes the form~\cite{Urrutia2021,Mukherjee2021}
\begin{equation}\label{eq3-1-2f}
R_{\rm BBH}(z_{\rm s})=R_0\int_{t_{\min}}^{t_{\max}}R_{\rm f}(t(z_{\rm s})-t_{\rm d})p(t_{\rm d}) dt_{\rm d},
\end{equation}
where $t_{\min}=50~{\rm Myr}$, $t_{\max}$ is set to the Hubble time, $P(t_{\rm d})\propto t_{\rm d}^{-1}$ is the time delay distribution which set the power law, and  $R_{\rm f}(t(z_{\rm s})-t_{\rm d})$ is the Star Formation Rate (SFR) as~\cite{Madau2016,Urrutia2021}
\begin{equation}\label{eq3-1-2}
R_{\rm f}(z_{\rm s})=[1+(1+z_{\rm p})^{-\lambda-\kappa}]\frac{(1+z_{\rm s})^{\lambda}}{1+\bigg(\frac{1+z_{\rm s}}{1+z_{\rm p}}\bigg)^{\lambda+\kappa}},
\end{equation}
where $\lambda=2.6$, $\kappa=3.6$, and $z_{\rm p}=2.2$. We use a power law mass distribution for the heavier black hole in the BBHs as~\cite{Liao2020,Basak2021,Urrutia2021,Zhou2022}
\begin{equation}\label{eq3-1-3}
P_{\rm BBH}(m_{\rm s})=Z_{\rm m}m_{\rm s}^{-\alpha}\mathcal{H}(m_{\rm s}-m_{\min})\mathcal{H}(m_{\max}-m_{\rm s}),
\end{equation}
where $m_s$ represents the mass of one of the celestial bodies in the binary star system, $\mathcal{H}$ is the Heaviside step function, and $Z_{\rm m}$ is the normalization constant as
\begin{equation}\label{eq3-1-4}
Z_{\rm m}=\left\{
\begin{aligned}
\frac{1-\alpha}{M_{\max}^{1-\alpha}-M_{\min}^{1-\alpha}},\quad\alpha\neq1,\\
\frac{1}{\ln(M_{\max}/M_{\min})},\quad\alpha=1.
\end{aligned}
\right.
\end{equation}
The heavier black hole $m_1$ satisfies the mass distribution as Eq.~(\ref{eq3-1-3}) with $\alpha=2.35$ in the range of $m_1\in[5, 100]~M_{\odot}$, and the mass of $m_2$ uniformly distributes in the interval $[m_1/{18},m_1]$.

\textbf{Generating the lensed GW events:} By assuming that the compact dark matter is distributed uniformly in comving volume, the probability that a GW source at $z_{\rm s}$ is lensed by an intervening compact object can be given by optical depth as
\begin{equation}\label{eq3-3-2}
P_{\rm L}(z_{\rm s})=1-\exp(-\tau(z_{\rm s})).
\end{equation}
We identify a BBH GW signal as a lensed event when $P_{\rm L}(z_{\rm s})$ is larger than a random number uniformly distribution between 0 and 1. Therein the optical depth $\tau(z_{\rm s})$ can be written as
\begin{equation}\label{eq3-3-3}
\tau(z_{\rm s})=\int_0^{+\infty}dm_{\rm len}\int_0^{z_{\rm s}}dz_{\rm l}\int_0^{y_0}dy~\tau(m_{\rm len},z_{\rm l},y),
\end{equation}
where $y_0$ is the maximum impact parameter, and $\tau(m_{\rm len}, z_{\rm l}, y)$ is the differential optical depth for the mass of lens $m_{\rm len}$, redshift of lens $z_{\rm l}$ and impact parameter $y$ which is given by
\begin{equation}\label{eq3-3-4}
\begin{split}
\tau(m_{\rm len}, z_{\rm l},  y)=&3yf_{\rm CO}\Omega_{\rm DM}\psi(m_{\rm len})\times\\
&\frac{H_0^2(1+z_{\rm l})^2}{H(z_{\rm l})}\frac{D_{\rm L}D_{\rm LS}}{D_{\rm S}}.
\end{split}
\end{equation}
where $f_{\rm CO}$ ($f_{\rm CO}\equiv\frac{\Omega_{\rm CO}}{\Omega_{\rm DM}}$) the fraction of compact dark matter in dark matter at present universe, $H(z_{\rm l})$ is the Hubble function at $z_{\rm l}$, and $\psi(m_{\rm len})$ is mass function for compact dark matter.

\textbf{Generating the parameters of lens:} When a GW event is identified as a lensed one, the probability distribution of  $z_{\rm l}$ an be obtained through the calculation of optical depth
\begin{equation}\label{eq3-3-5}
P(z_{\rm l})=\frac{\tau(m_{\rm len},z_{\rm l},y)}{\int_0^{z_{\rm s}}dz_{\rm l}~\tau(m_{\rm len},z_{\rm l},y)}.
\end{equation}
In addition, the distribution $P(y)$ also can be obtained from Eq.~(\ref{eq3-3-4}) in the range of $y\in[0,y_0]$
\begin{equation}\label{eq3-3-6}
P(y)=\frac{\tau(m_{\rm len},z_{\rm l},y)}{\int_0^{y_0}dy~\tau(m_{\rm len},z_{\rm l},y)}=\frac{2}{y_0^2}y,
\end{equation}
we adopt $y_0=3$ as the cutoff value since it becomes difficult to confidently identify a GW event as lensed when  $y$ exceeds this threshold.  For the lens mass distribution $\psi(m_{\rm len})$, we assume a uniform distribution within the minimum value of the mass function $M_{\min}=1~M_{\odot}$ and the maximum value $M_{\max}=10^3~M_{\odot}$.  For a more detailed description of the calculation of compact dark matter distribution, please refer to  references \cite{Shan:2023ngi,Lin:2025mpx}.

As shown in Eq. (\ref{eq5}), the amplification factor depends on two key parameters: (i) the redshifted lens mass $M_{\rm Lz}=m_{\rm len}(1+z_{\rm l})$ and (ii)  impact parameter $y$. The influence of these parameters on gravitational waves has been extensively studied in previous works \cite{Takahashi_2003, Lin2023, Suyamprakasam:2025pvk, Liao:2019aqq}. However, most existing studies assume uniform distributions for both $M_{\rm Lz}$ and $y$,which is astrophysically unrealistic. First, $M_{\rm Lz}$ incorporates the lens redshift $z_{\rm l}$,  which follows the intrinsic distribution of astrophysical objects. Second, $y$ represents the source–lens offset; smaller values of $y$ correspond to higher lensing probabilities. In this work, we adopt physically motivated distributions for both parameters to better reflect realistic astrophysical conditions. This constitutes one of the key contributions of our paper. It should be emphasized, however, that while these distributions are chosen for physical consistency, our gravitational wave lensing identification algorithm remains fully general and applicable to arbitrary mass distributions and parameter choices.

\subsection{Whitening Preprocessing}
Gravitational wave detectors, such as LIGO and Virgo, exhibit non-uniform noise characteristics, with varying intensity across different frequencies and typically greater noise at extreme frequencies. This non-uniform noise can obscure genuine signal features.

The whitening process aims to eliminate this frequency-dependent noise. Ideally, whitening transforms the noise into white noise, which has uniform variance across all frequencies. This allows genuine signal features to emerge more prominently in time--frequency representations without distortion from the detector's noise spectrum.

Whitening is performed in the frequency domain. For a discrete time series $ x[n] $ with Fourier transform $ X[k] $ (where $ k $ denotes the frequency index), the whitened Fourier coefficients are given by:
\begin{equation}
X_{\text{white}}[k] = \frac{X[k]}{\sqrt{S_n(f_k)}},
\end{equation}
where $f_k$ is the frequency corresponding to index $ k $, and $ \sqrt{S_n(f_k)} $ represents the standard deviation of the noise at that frequency, derived from the noise power spectral density (PSD) $ S_n(f_k) $. This normalization effectively standardizes the noise to unit variance. Whitening is typically applied to the raw time series data $ x[n] $ before subsequent processing steps such as the Q-transform.

\subsection{Q-Transform Description}

The Q-transform technique has long been used as a visualization tool in the LIGO-Virgo data analysis. It is a modified short-time Fourier transform where the analysis window duration varies inversely with frequency such that the plane of time-frequency is covered by the tiles of constant $Q$. The continuous Q-transform is given by
\begin{equation}
x(\tau,f)=\int_{-\infty}^{\infty}x(t)\:w(t-\tau)\:\exp(-2\pi ift)\:\mathrm{d}t,
\end{equation}
where $w(t-\tau,f)$ is a window function centered at $\tau$ and depends on the quality factor $Q$,
\begin{equation}
Q=\frac{f}{\delta f}.
\end{equation}
The discrete version of the above is more appropriate for gravitational wave data, as this data is collected in discrete time bins,
\begin{equation}
x[m,k]=\sum_{n=0}^{N-1}x[n]w[n-m,k]\exp\left(-2\pi ink/N\right).
\end{equation}
When we perform Q-transform of a signal, we have to specify a range of values of $Q$. The value of $Q$ which corresponds to the largest SNR is chosen. This transformation creates a tile in time and frequency domain and the value in each tile corresponds to the energy of the signal.

\begin{figure*}[!t]
  \centering
  \includegraphics[width=1\linewidth]{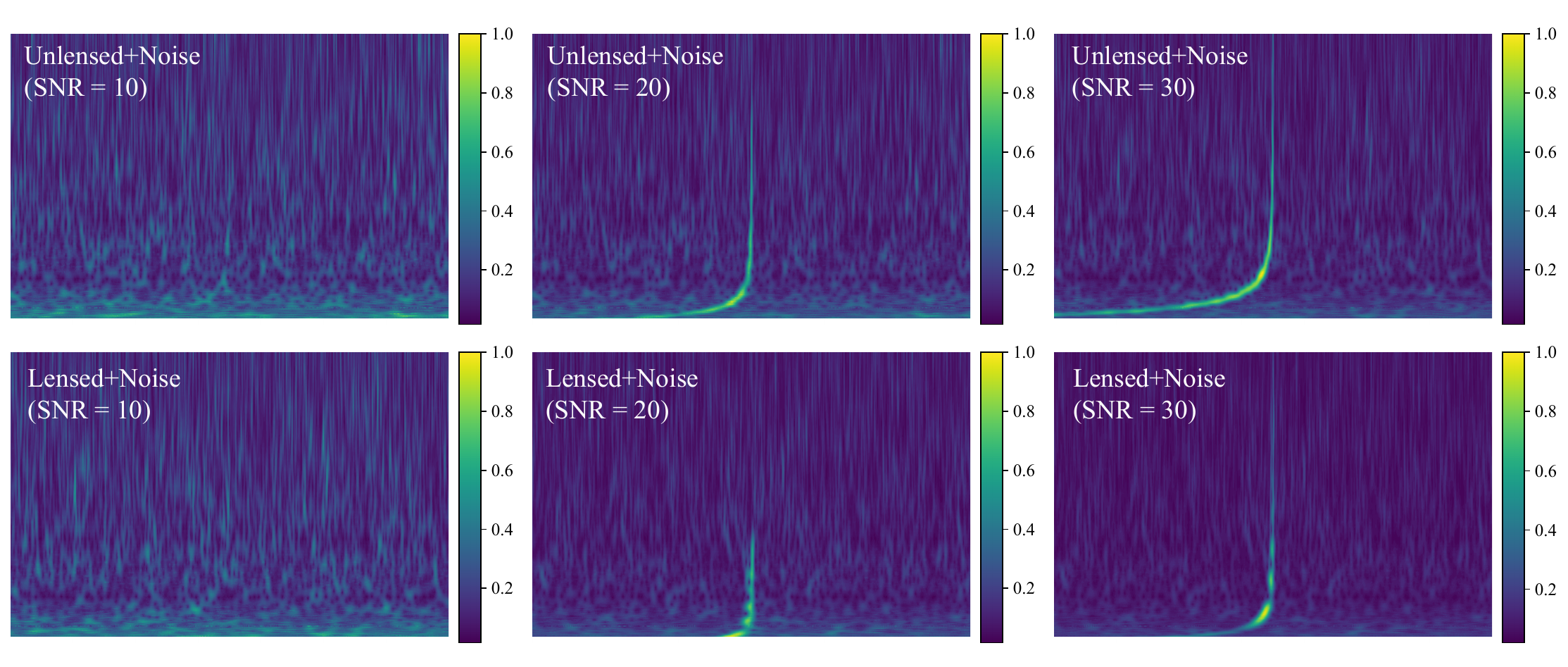}
  \caption{Comparison of time-frequency images between unlensed  and lensed  gravitational wave signals across varying noise levels (SNR = 10, 20, 30).}
  \label{fig:spectral_plots}
\end{figure*}

\section{Machine Learning}

\subsection{Wavelet Convolutional Detector}
CNNs represent a class of deep feedforward neural networks characterized by convolutional operations~\cite{lecun2015deep, krizhevsky2012imagenet, dong2015image, szegedy2015going}. Their primary advantages stem from the utilization of \textit{local connectivity} and \textit{weight sharing}, which reduce the number of parameters, thereby facilitating network optimization and mitigating model complexity and the risk of overfitting. These advantages become particularly pronounced when processing image inputs. However, compared to Vision Transformers (ViTs) that perform global information aggregation~\cite{vit2021}, CNNs inherently rely on local convolution and pooling operations. This intrinsic bias towards encoding local patterns may limit their performance when handling complex inputs such as gravitational wave time-frequency images. While increasing the kernel size can expand the receptive field, studies indicate that this approach rapidly reaches a plateau and saturates.

\begin{figure*}[!t]
  \centering
  \includegraphics[width=0.9\textwidth]{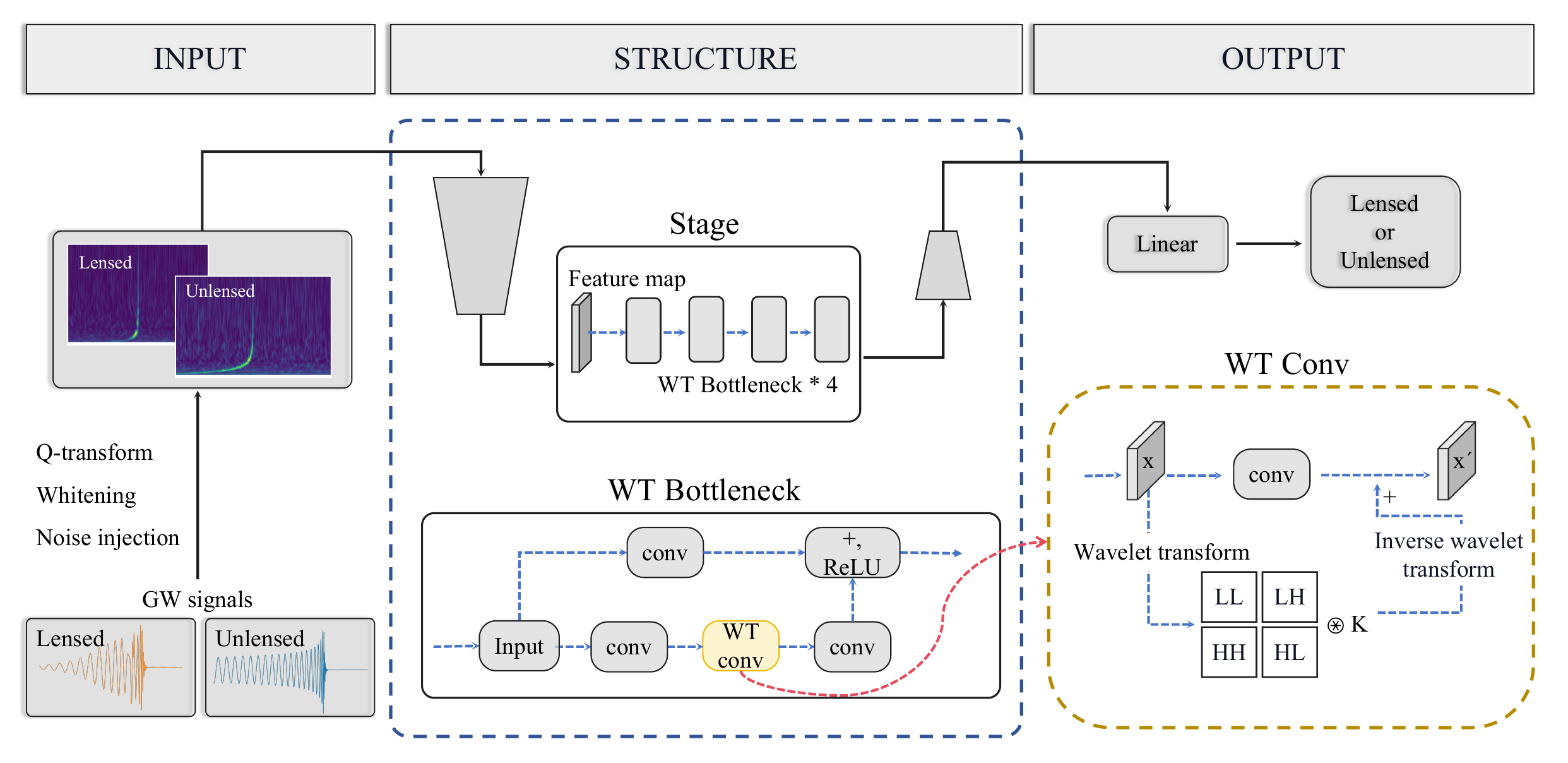}
  \caption{Architecture of the Wavelet Convolutional Detector featuring WTConv modules integrated within residual blocks.
  The wavelet-transform convolution layers enable multi-scale feature extraction critical for gravitational wave analysis.}
  \label{fig:model_arch}
\end{figure*}

There is research proving that wavelet transform can obtain a larger receptive field without causing excessive parameterization~\cite{finder2024wavelet}. Therefore, adopting ResNet as our baseline model~\cite{he2016deep}, we integrated Wavelet Transform Convolution (WTConv) layers within the network blocks to enhance model performance. The detailed architecture is illustrated in Figure~\ref{fig:model_arch}.

\subsubsection{Architecture Overview}
The network input consists of spectrograms generated by applying noise whitening (based on the PSD) and a Q-transform to the raw gravitational-wave (GW) signals. The proposed network adopts a ResNet backbone with five hierarchical stages, where conventional bottleneck blocks are augmented with Wavelet Transform Convolution (WTConv) modules. The architecture processes input tensors through sequential stages: Stage 0 performs initial feature extraction using standard convolutions, while Stages 1-4 employ wavelet-enhanced bottlenecks. The fundamental transformation across all stages follows the residual learning principle:
\begin{equation}
X_{\text{out}} = \mathcal{F}(X_{\text{in}}) + X_{\text{in}},
\end{equation}
where $\mathcal{F}$ represents the wavelet-enhanced residual function. The key innovation resides in the WTConv module that replaces standard convolutions, enabling multi-scale feature extraction critical for gravitational wave analysis.

The network gradually transforms features through four wavelet enhancement stages, as shown in Table \ref{table:stage}.
\vspace{-0.5cm}
\begin{table}[htbp!]
\renewcommand\arraystretch{1.5}
\centering
\caption{Wavelet Enhancement Stages}
\begin{tabular}{c c c}
\hline \hline
\textbf{Stage} & \textbf{Input Size} & \textbf{Output Size} \\ \hline
Stage 1 & $64\times64\times64$ & $256\times64\times64$ \\
Stage 2 & $256\times64\times64$ & $512\times32\times32 $\\
Stage 3 & $512\times32\times32$ & $1024\times16\times16$ \\
Stage 4 & $1024\times16\times16$ & $2048\times8\times8 $ \\
\hline\hline
\end{tabular}
\label{table:stage}
\end{table}

Each stage takes the output from the previous stage as input and produces a new output with increased depth and reduced spatial dimensions. This progressive transformation allows the network to capture features at multiple scales and levels of abstraction.

Spatial downsampling occurs at stage transitions via stride-2 WT-Bottleneck operations, preserving high-frequency components better than conventional pooling. The final classification head consists of global average pooling followed by a fully connected layer, implementing the complete transformation:
\begin{equation}
f(X) = \text{FC} \circ \Phi_{\text{stage4}} \circ \cdots \circ \Phi_{\text{stage0}}(X),
\end{equation}
where $\Phi_{\text{stage}}$ denotes wavelet-enhanced residual blocks.

\subsubsection{WTConv Operational Details}
The WTConv module implements frequency-adaptive processing through dual-path computation:
\begin{align}\nonumber
&\mathcal{H}(X) =\text{Conv}_{k\times k}(X), \\
&\Omega(X) =\mathcal{W}^{-1}\left(\mathbf{K}\circledast\mathcal{W}(X)\right), \\ \nonumber
&\text{WTConv}(X) =\mathcal{H}(X)+\Omega(X) .\nonumber
\end{align}
The wavelet transform $\mathcal{W}$ decomposes features into four sub-bands:
\begin{equation}
\mathcal{W}(X) = Z = [\text{LL}, \text{LH}, \text{HL}, \text{HH}] \in \mathbb{R}^{C \times 4 \times H/2 \times W/2},
\end{equation}
where $Z$ denotes the concatenated low–pass (LL) and three high–pass (LH, HL, HH) coefficient tensors obtained after decomposition. The inverse transform $\mathcal{W}^{-1}$ reconstructs features through transposed convolution:
\begin{equation}
\mathcal{W}^{-1}(Z) = \text{ConvTranspose}_{2\times2}(Z) \in \mathbb{R}^{C \times H \times W}.
\end{equation}

Between transforms, wavelet-domain processing applies depthwise convolution and channel-wise scaling:
\begin{equation}
Z_{\text{processed}} = \mathbf{S} \odot (\mathbf{K} \circledast Z),
\end{equation}
where $\mathbf{K}\in\mathbb{R}^{C\times 4\times k\times k}$ is the learnable depthwise kernel acting on the four sub-bands of $Z$, and $\mathbf{S}\in\mathbb{R}^{C\times 4}$ denotes channel-wise learnable scaling parameters; $\odot$ represents element-wise multiplication. This lightweight processing captures cross-band dependencies while maintaining computational efficiency.

WTConv's frequency-adaptive processing proves particularly advantageous at this mid-level hierarchy, where gravitational-wave signatures exhibit distinctive time-frequency patterns.

The network culminates with global average pooling reducing feature maps to $2048 \times 1 \times 1$, followed by a batch normalization layer, ReLU activation, and a final fully-connected layer projecting features to binary classification outputs.

The WTConv module fundamentally enhances the network's ability to capture multi-scale signatures in gravitational wave data through three mechanisms: 1) Expanded receptive fields without kernel size inflation, where a single-level WTConv achieves effective $6\times6$ coverage versus $3\times3$ in standard convolution; 2) Built-in frequency separation that processes low-frequency structures and high-frequency transients in parallel subbands; 3) Parameter efficiency through depthwise wavelet processing, offering marked reductions in parameter count compared to conventional convolutions with equivalent spatial coverage. These properties make WTConv particularly suitable for gravitational-wave analysis, where signal morphology manifests as localized time-frequency patterns embedded in non-stationary noise backgrounds.

\subsection{Data Generation Pipeline}

The gravitational wave data generation initiates with waveform synthesis, we  employ Python module  \textsf{PyCBC} to produce time-domain signals $h(t)$ for binary black hole systems~\cite{alex}. Utilizing the \textsf{IMRPhenomPv2} waveform model, this process incorporates critical physical parameters including component masses and luminosity distance, with appropriate low-frequency cutoff settings to match detector characteristics. Subsequently, gravitational lensing effects are introduced through Fourier-domain operations: the original waveform is transformed to the frequency domain $\widetilde{h}(f)$, modulated by a diffraction transfer function $F(f)$ to yield the lensed spectrum $\widetilde{h}^{\rm L}(f) = F(f) \cdot \widetilde{h}(f)$, and then inverse-transformed to reconstruct the lensed time-series $\widetilde{h}^{\rm L}(f)$.

To simulate realistic observational conditions, colored noise $n(t)$ conforming to ET's power spectral density $S_n(f)$ is generated and additively combined with the lensed waveform, resulting in the noisy signal $s(t) = \widetilde{h}^{\rm L}(f) + n(t)$.
Whitening processing is then applied to enhance signal detectability: $s(t)$ undergoes Fourier transformation followed by spectral normalization via division by $\sqrt{S_n(f)}$ in the frequency domain, with careful handling of power spectrum interpolation and boundary artifact suppression. The inverse transform of the normalized spectrum produces the whitened signal $s_{\rm white}(t)$. The lensed GW signals are embedded within the noisy background with a SNR $\in$  [10, 50].

The final stage involves time-frequency characterization through continuous wavelet transformation using Morlet wavelets. The magnitude spectrum $A(t,f) = |Q(t,f)|$ is computed from the wavelet coefficients, and the characteristic spectrogram is generated by visualizing $\log_{10}A(t,f)$ across the time-frequency plane. Comprehensive parameter ranges governing all physical and numerical aspects of this pipeline are systematically documented in Table \ref{tab:parameters}, while representative output visualizations are presented in Fig. \ref{fig:spectral_plots}.

\begin{table*}[!t]
\centering
\renewcommand\arraystretch{1.3}
\caption{Parameter ranges used in lensed gravitational wave simulation}
\label{tab:parameters}
\setlength{\tabcolsep}{12pt} 
\begin{tabular*}{\textwidth}{@{\extracolsep{\fill}} l l l l l@{}}
\toprule
\hline \hline
\textbf{Parameter} & \textbf{Description} & \textbf{Shape} &\textbf{Range} & \textbf{Unit} \\
\hline
\multicolumn{4}{c}{\emph{Binary System Parameters}} \\
\hline
$m_1$ &  One of mass of the BH & Distribution from Eq. (\ref{eq3-1-3}) & [5, 100] & $M_\odot$ \\
$m_2$ & One of mass of the BH & Same with $m_1$ &$ [m_1/18, m_1]$ & $M_\odot$ \\
$a_1$ &   Dimensionless spin of each BH & Uniform & $ [0, 0.99]$ & --\\
$a_2$  & Dimensionless spin of each BH & Uniform & $ [0, 0.99]$ & --\\
$\iota$ & Inclination angle & Sine & [0, $\pi$] & rad \\
$\phi_{\rm c}$ & Phase at the moment of coalescence & Uniform & [0, $2\pi$] & rad \\
$\psi$ & Polarization angle &  Uniform &  [0, $\pi$] & rad \\
$z_{\rm s}$ & Source Redshift &  Distribution from  Eq. (\ref{eq3-1-1})  & -- & -- \\
$ra$ & Ecliptic longitude of the binary system & Uniform &  [0, $2\pi$] & rad \\
$dec$ & Ecliptic latitude of the binary system & Cos & [-$\pi/2$,$\pi/2$] & rad \\
$\delta t$ & The difference of geocent time & Uniform & [0, $3600\times24\times30$] & s \\
\midrule
\hline
\multicolumn{4}{c}{\emph{Lensing Parameters}} \\
\hline
\midrule
$z_{\rm l}$ &  Lens Redshift  & Distribution from Eq. (\ref{eq3-3-5}) &-- & -- \\
$y$ & Impact parameter & Distribution from Eq. (\ref{eq3-3-6}) & [0, 3] & -- \\
$m_{\rm len}$ & Mass of the lens & Uniform & [1, 1000] & $M_\odot$ \\
\bottomrule
\hline \hline
\end{tabular*}
\end{table*}

Time-domain strain series are synthesized by injecting GW signals into Gaussian noise.
We assemble a background population of BBH mergers and model their lensing by massive early-type galaxies via a point-mass lens model.
The resulting catalogue comprises 5000 lensed and 5000 unlensed events, partitioned into training, validation, and test subsets in the ratio 8 : 1 : 1.
The test subset is further split into high-SNR and low-SNR segments of equal cardinality.

\subsection{Hyperparameter Setting}
The hyperparameter settings are shown in Table \ref{table: hyper-parameters}, we fix the training and testing batch size at 32 and resize all input images to 256\,$\times$\,256 pixels.
The maximum number of training epochs is set to 500.
To accelerate convergence and enhance model stability, we employ the Adam optimiser \cite{Kingma2014AdamAM} with an initial learning rate of $1\times10^{-3}$.
A cosine-annealing scheduler is adopted \cite{sgdr2017}, with $T_{\text{max}}=500$, driving the learning rate smoothly to zero.
For reproducibility, the random seed is fixed at 42.
All experiments are conducted on an NVIDIA GeForce RTX~3090 (24\,GB RAM) using CUDA~11.7 and PyTorch~2.0.1.
At the end of each epoch, we evaluate the loss and accuracy on the validation set and save the best checkpoint.

\vspace{-0.5cm}
\begin{table}[htbp]
\centering
\renewcommand\arraystretch{1.3}
\caption{Summary of hyper-parameters.}
\begin{tabular}{ll}
\hline \hline
\textbf{Hyper-parameter} & \textbf{Value} \\ \hline
Batch size               & 32 \\
Input size               & $256\times256$ \\
Max epochs               & 500 \\
Optimiser                & Adam \\
Initial learning rate    & $1\times10^{-3}$ \\
Scheduler                & Cosine annealing ($T_{\text{max}}=500$, $\eta_{\text{min}}=0$) \\
Random seed              & 42 \\
GPU                      & RTX~3090 (24\,GB) \\
Software                 & CUDA~11.7 + PyTorch~2.0.1 \\ \hline \hline
\end{tabular}
\label{table: hyper-parameters}
\end{table}

\section{Results and discussions}
\subsection{Model training process}

The training dynamics of the model demonstrated robust convergence and high performance across all evaluated metrics. The training loss decreased sharply in the early phases of training and stabilized at approximately $8.0 \times 10^{-5}$, reflecting a well-regularized training process. On the validation set, the model achieved an accuracy of $9.2 \times 10^{-1}$ and an AUC value of $9.5 \times 10^{-1}$, highlighting its strong discriminative capability and reliability in classifying unseen data. The learning rate was methodically reduced from $1.0 \times 10^{-3}$ to $1.0 \times 10^{-6}$ following an exponential decay schedule, which enabled refined parameter updates and steady convergence toward optimal solutions. These results collectively affirm the model's exceptional generalization performance and the effectiveness of the proposed training framework.

\subsection{Confusion Matrix}
\begin{figure*}[htbp]
  \centering
  {\includegraphics[width=0.48\linewidth]{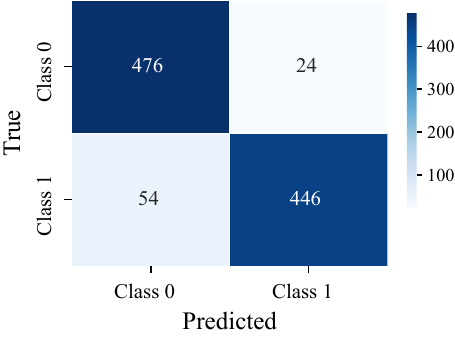}
  \includegraphics[width=0.48\linewidth]{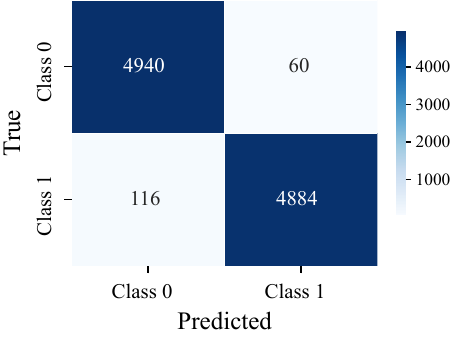}}
  \caption{\textit{Left}: Confusion matrix for the validation dataset; \textit{Right}: Confusion matrix for the full dataset.}
  \label{fig:confusion_matrices}
\end{figure*}

The confusion matrix presented in Figure \ref{fig:confusion_matrices}  offers a comprehensive summary of the model’s classification performance across the entire datasets of  validation and full  gravitational wave images. Class 0 corresponds to unlensed gravitational waves, while Class 1 represents lensed gravitational waves.

Overall, the model achieves exceptionally high accuracy in distinguishing between lensed and unlensed signals. It correctly classified 4,940 instances of unlensed gravitational waves and 4,884 instances of lensed waves. The number of misclassifications was notably low: only 60 unlensed signals were incorrectly identified as lensed (false positives), and 116 lensed signals were misclassified as unlensed (false negatives), resulting in an overall accuracy of 98.24\%. The model exhibits strong generalization capability across the dataset with remarkably low false positive and false negative rates. A slight asymmetry in errors is observed, with a marginally higher number of lensed signals being misclassified as unlensed, which may reflect inherent challenges in recognizing certain lensed waveform morphologies—though this discrepancy remains modest.

On the validation set, the model also demonstrates robust performance, as reflected in right panel of Figure \ref{fig:confusion_matrices} left:
This corresponds to 476 correctly classified unlensed samples and 446 correctly classified lensed samples, with 24 false positives and 54 false negatives, yielding a validation accuracy of 92.2\%. Although slightly lower than the full-set accuracy, this result remains high and consistent with the overall trend. specifically, the model continues to show a slight tendency to misclassify lensed signals as unlensed. The stability in performance across both datasets confirms the model’s strong discriminative capability without evident overfitting. These outcomes underscore the effectiveness of the proposed approach in capturing complex features associated with gravitational wave lensing, supporting its potential applicability in large-scale data analysis.

\subsection{Receiver Operating Characteristic curves}

To systematically investigate model behavior under varying physical conditions, we partition the test sets based on three critical parameters: SNR, impact parameter $y$, and lens mass $m_{\rm len}$. Each subset is evaluated using Receiver Operating Characteristic (ROC) curves, which plot the true positive rate (TPR) against the false positive rate (FPR) across varying classification thresholds~\citep{k33}. The area under the ROC curve (AUC) serves as a comprehensive metric for overall discriminative capability, with higher values indicating superior classification performance.

To mitigate bias from uneven subset sizes and ensure statistical robustness of performance comparisons, we establish the following physically motivated stratification thresholds: Samples with SNR $>$ 30 are considered high-SNR, others are low-SNR; The threshold of impact parameter $y$  is 2; Samples with $m_{\rm len} > 500 M_\odot$ are high lens mass samples, others are low.

\begin{figure*}[htbp]
\centering
\includegraphics[width=1\linewidth]{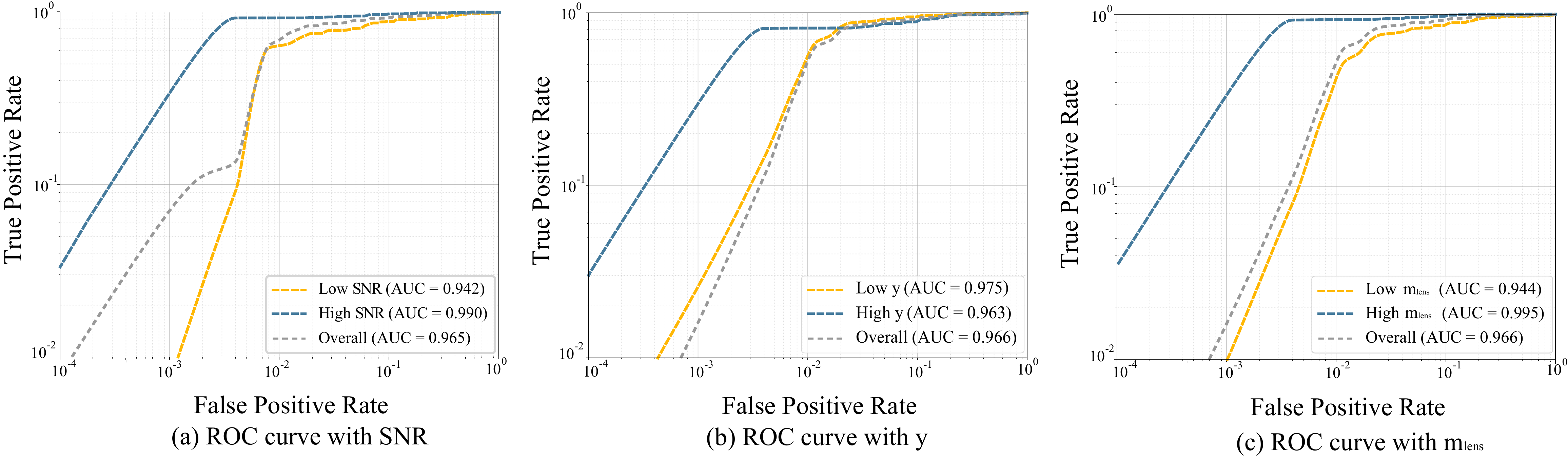}
\caption{ROC curves evaluating model performance:
(a) SNR dependence: Model performance improves with increasing SNR;
(b) Impact parameter $y$ dependence: Systematic AUC enhancement with decreasing $y$ due to amplified wave-optics effects in strong-field regions;
(c) Lens mass $m_{\rm len}$ dependence: Performance gains with increasing mass resulting from intensified lensing artifacts in deeper gravitational potentials.}
\label{fig:ROC}
\end{figure*}

The ROC curves depicted in Figure \ref{fig:ROC} provide a comprehensive evaluation of the model's ability to distinguish between unlensed and lensed gravitational waves under different physical conditions. The overall ROC curve (AUC = 0.965) indicates strong classification performance across the entire dataset. The curve rises sharply from the origin, demonstrating a high true positive rate even at low false positive rates, which underscores the model's general effectiveness.

For low-SNR subsets (AUC = 0.942), the modest performance decline stems from noise-induced degradation of high-frequency waveform components—critical for identifying lensing artifacts. Gravitational wave signals from compact binary mergers (the primary source of lensed signals in this study) exhibit most lensing-related distortions in the  frequency range. In low-SNR scenarios,  noise masks these high-frequency features, reducing the model's ability to distinguish between unlensed signals (which have smooth, single-peaked waveforms) and lensed signals (which have modulated, multi-peaked waveforms). This is evident in the ROC curve's shallower slope at low FPR: for FPR=0.01, the TPR drops to $\sim$0.78 (vs. $\sim$0.85 in the overall dataset), reflecting increased ambiguity between noise-corrupted lensed signals and clean unlensed signals. In contrast, the high-SNR subset (AUC=0.990) achieves near-perfect performance, with the ROC curve approaching the ideal top-left corner (TPR $\approx$1 at FPR$<$0.001)—high SNR ensures lensing-induced waveform modifications are clearly resolved against noise, allowing the model to exploit these high-fidelity features for confident classifications, minimizing false negatives and false positives.

The model's discriminative ability exhibits a strong monotonic dependence on the fundamental lensing parameters $y$, with AUC increasing systematically as $y$ decreases. This trend is not arbitrary but directly reflects the underlying gravitational wave lensing physics:
The impact parameter $y$  quantifies the wave proximity to the lens's gravitational potential: smaller
$y$ means the  wave passes closer to the lens center, entering the strong-field regime where spacetime curvature is most extreme. For $y<1$, the  wave generates significant distinct diffraction effect and localized amplitude enhancements. These effects introduce sharp, reproducible features in the waveform that the model can easily learn. In contrast, for $y>2$, the  wave only experiences weak  waveform distortions (e.g., about 5\% change in peak amplitude) that are difficult to distinguish from intrinsic variations in unlensed signals (e.g., variations in binary mass ratio). This explains the observed AUC decline with increasing $y$, the AUC drops to ~0.88 for low $y$ subsets (vs.~0.97 for high $y$ subsets), as the model lacks sufficient discriminative features to separate weakly lensed and unlensed signals.

For lens mass $m_{\rm len}$—the key factor governing gravitational wave diffraction (the primary lensing signature in our analysis)—its influence stems from the scaling of the Einstein radius $R_{\rm E} \propto \sqrt{m_{\rm len}}$, which determines the detectability of diffraction effects and thus the model’s performance: for high-$m_{\rm len}$ samples ($m_{\rm len}>500 \, M_\odot$), the larger $R_{\rm E}$ matches the gravitational wave wavelength $\lambda_{\rm GW}$ ($\propto 1/f$, $f=1\text{--}1000 \, \text{Hz}$ for compact binaries), reproducible features the model leverages for high performance (AUC=0.995); in contrast, low-$m_{\rm len}$ samples ($m_{\rm len}\leq500\, M_\odot$) have $R_{\rm E}\ll\lambda_{\rm GW}$, suppressing diffraction to minimal, noise-like waveform perturbations that are masked by detector noise or confused with intrinsic signal variations, leading to degraded performance (AUC=0.944). This trend aligns with wave-optics theory (diffraction prominence requires $R_{\rm E} \sim \lambda_{\rm GW}$), validating the model decodes genuine diffraction-induced lensing imprints rather than spurious correlations.

\section{Conclusion}
This work  first presents the machine learning-based method for identifying wave-optics  signatures in gravitationally lensed gravitational waves. We propose the WCD, a novel deep learning framework specifically designed to capture the intricate diffraction patterns produced by microlensing from compact dark matter. By directly targeting these wave-optics effects—which are typically overlooked in traditional geometric-optics analyses—the WCD establishes a new state-of-the-art in distinguishing lensed gravitational wave events. The key contributions of this study include:

(i) Diffraction-aware feature extraction through the integration of multi-scale wavelet transforms within residual convolutional blocks. This architecture efficiently captures intricate time-frequency interference patterns induced by wave-optics effects, providing a larger effective receptive field than standard convolutions while maintaining high parameter efficiency.

(ii) Astrophysically consistent training data generation that incorporates realistic compact DM mass distributions and redshift-dependent lensing probabilities. This ensures the model learns physically meaningful features rooted in actual lensing environments, overcoming the limitations of uniform parameter sampling used in previous studies.

(iii) Machine learning-based adaptability to a wide range of lens masses. Unlike traditional matched filtering techniques that require precise template banks for specific lens masses, the WCD demonstrates robust performance across a broad spectrum of lens masses—from stellar-mass compact objects to intermediate-mass dark matter without retraining or template adjustments.

Experimental results show that the WCD achieves an overall accuracy of 92\% and an AUC of 0.965 on a synthetic dataset of 10,000 BBH events. The model performs exceptionally well under high-SNR conditions (AUC $\sim$ 0.99) and exhibits physically consistent behavior, with performance improving as the impact parameter decreases and the lens mass increases. Most notably, the model maintains high discriminative power across the entire range of lens masses, highlighting its suitability for detecting lensing by compact DM objects whose mass distribution remains poorly constrained.

With inference speeds orders of magnitude faster than Bayesian methods, the WCD offers a scalable and efficient solution for large-scale lensing searches in the era of third-generation gravitational wave detectors. This framework not enhances the potential for discovering and characterizing dark matter substructures through gravitational wave lensing but also provides a flexible and robust tool for probing wave-optics effects across diverse lens populations.

\section*{Acknowledgments}
This work was supported by National Key R$\&$D Program of China (No. 2024YFC2207400), the National Natural Science Foundation of China under Grants No. 12203009, 12374408  and 12475051 and the science and technology innovation Program of Hunan Province under grant No. 2024RC1050.


\bibliography{ref}

\begin{thebibliography}{70}
\expandafter\ifx\csname natexlab\endcsname\relax\def\natexlab#1{#1}\fi
\expandafter\ifx\csname bibnamefont\endcsname\relax
  \def\bibnamefont#1{#1}\fi
\expandafter\ifx\csname bibfnamefont\endcsname\relax
  \def\bibfnamefont#1{#1}\fi
\expandafter\ifx\csname citenamefont\endcsname\relax
  \def\citenamefont#1{#1}\fi
\expandafter\ifx\csname url\endcsname\relax
  \def\url#1{\texttt{#1}}\fi
\expandafter\ifx\csname urlprefix\endcsname\relax\def\urlprefix{URL }\fi
\providecommand{\bibinfo}[2]{#2}
\providecommand{\eprint}[2][]{\url{#2}}

\bibitem[{\citenamefont{{Abbott}
  et~al.}(2016{\natexlab{a}})\citenamefont{{Abbott}, {Abbott}, {Abbott},
  {Abernathy}, {Acernese}, {Ackley}, {Adams}, {Adams}, {Addesso}, {Adhikari}
  et~al.}}]{PhysRevLett.116.061102}
\bibinfo{author}{\bibfnamefont{B.~P.} \bibnamefont{{Abbott}}},
  \bibinfo{author}{\bibfnamefont{R.}~\bibnamefont{{Abbott}}},
  \bibinfo{author}{\bibfnamefont{T.~D.} \bibnamefont{{Abbott}}},
  \bibinfo{author}{\bibfnamefont{M.~R.} \bibnamefont{{Abernathy}}},
  \bibinfo{author}{\bibfnamefont{F.}~\bibnamefont{{Acernese}}},
  \bibinfo{author}{\bibfnamefont{K.}~\bibnamefont{{Ackley}}},
  \bibinfo{author}{\bibfnamefont{C.}~\bibnamefont{{Adams}}},
  \bibinfo{author}{\bibfnamefont{T.}~\bibnamefont{{Adams}}},
  \bibinfo{author}{\bibfnamefont{P.}~\bibnamefont{{Addesso}}},
  \bibinfo{author}{\bibfnamefont{R.~X.} \bibnamefont{{Adhikari}}},
  \bibnamefont{et~al.}, \bibinfo{journal}{\prl} \textbf{\bibinfo{volume}{116}},
  \bibinfo{eid}{061102} (\bibinfo{year}{2016}{\natexlab{a}}),
  \eprint{1602.03837}.

\bibitem[{\citenamefont{{Abbott}
  et~al.}(2016{\natexlab{b}})\citenamefont{{Abbott}, {Abbott}, {Abbott},
  {Abernathy}, {Acernese}, {Ackley}, {Adams}, {Adams}, {Addesso}, {Adhikari}
  et~al.}}]{k38}
\bibinfo{author}{\bibfnamefont{B.~P.} \bibnamefont{{Abbott}}},
  \bibinfo{author}{\bibfnamefont{R.}~\bibnamefont{{Abbott}}},
  \bibinfo{author}{\bibfnamefont{T.~D.} \bibnamefont{{Abbott}}},
  \bibinfo{author}{\bibfnamefont{M.~R.} \bibnamefont{{Abernathy}}},
  \bibinfo{author}{\bibfnamefont{F.}~\bibnamefont{{Acernese}}},
  \bibinfo{author}{\bibfnamefont{K.}~\bibnamefont{{Ackley}}},
  \bibinfo{author}{\bibfnamefont{C.}~\bibnamefont{{Adams}}},
  \bibinfo{author}{\bibfnamefont{T.}~\bibnamefont{{Adams}}},
  \bibinfo{author}{\bibfnamefont{P.}~\bibnamefont{{Addesso}}},
  \bibinfo{author}{\bibfnamefont{R.~X.} \bibnamefont{{Adhikari}}},
  \bibnamefont{et~al.}, \bibinfo{journal}{\apjs}
  \textbf{\bibinfo{volume}{225}}, \bibinfo{eid}{8}
  (\bibinfo{year}{2016}{\natexlab{b}}), \eprint{1604.07864}.

\bibitem[{\citenamefont{{Abbott}
  et~al.}(2019{\natexlab{a}})\citenamefont{{Abbott}, {Abbott}, {Abbott},
  {Abraham}, {Acernese}, {Ackley}, {Adams}, {Adhikari}, {Adya}, {Affeldt}
  et~al.}}]{k2}
\bibinfo{author}{\bibfnamefont{B.~P.} \bibnamefont{{Abbott}}},
  \bibinfo{author}{\bibfnamefont{R.}~\bibnamefont{{Abbott}}},
  \bibinfo{author}{\bibfnamefont{T.~D.} \bibnamefont{{Abbott}}},
  \bibinfo{author}{\bibfnamefont{S.}~\bibnamefont{{Abraham}}},
  \bibinfo{author}{\bibfnamefont{F.}~\bibnamefont{{Acernese}}},
  \bibinfo{author}{\bibfnamefont{K.}~\bibnamefont{{Ackley}}},
  \bibinfo{author}{\bibfnamefont{C.}~\bibnamefont{{Adams}}},
  \bibinfo{author}{\bibfnamefont{R.~X.} \bibnamefont{{Adhikari}}},
  \bibinfo{author}{\bibfnamefont{V.~B.} \bibnamefont{{Adya}}},
  \bibinfo{author}{\bibfnamefont{C.}~\bibnamefont{{Affeldt}}},
  \bibnamefont{et~al.}, \bibinfo{journal}{\prd} \textbf{\bibinfo{volume}{100}},
  \bibinfo{eid}{104036} (\bibinfo{year}{2019}{\natexlab{a}}),
  \eprint{1903.04467}.

\bibitem[{\citenamefont{{Berti} et~al.}(2011)\citenamefont{{Berti}, {Gair}, and
  {Sesana}}}]{k39}
\bibinfo{author}{\bibfnamefont{E.}~\bibnamefont{{Berti}}},
  \bibinfo{author}{\bibfnamefont{J.}~\bibnamefont{{Gair}}}, \bibnamefont{and}
  \bibinfo{author}{\bibfnamefont{A.}~\bibnamefont{{Sesana}}},
  \bibinfo{journal}{\prd} \textbf{\bibinfo{volume}{84}}, \bibinfo{eid}{101501}
  (\bibinfo{year}{2011}), \eprint{1107.3528}.

\bibitem[{\citenamefont{{Acernese} et~al.}(2015)\citenamefont{{Acernese},
  {Agathos}, {Agatsuma}, {Aisa}, {Allemandou}, {Allocca}, {Amarni}, {Astone},
  {Balestri}, {Ballardin} et~al.}}]{Acernese_2015}
\bibinfo{author}{\bibfnamefont{F.}~\bibnamefont{{Acernese}}},
  \bibinfo{author}{\bibfnamefont{M.}~\bibnamefont{{Agathos}}},
  \bibinfo{author}{\bibfnamefont{K.}~\bibnamefont{{Agatsuma}}},
  \bibinfo{author}{\bibfnamefont{D.}~\bibnamefont{{Aisa}}},
  \bibinfo{author}{\bibfnamefont{N.}~\bibnamefont{{Allemandou}}},
  \bibinfo{author}{\bibfnamefont{A.}~\bibnamefont{{Allocca}}},
  \bibinfo{author}{\bibfnamefont{J.}~\bibnamefont{{Amarni}}},
  \bibinfo{author}{\bibfnamefont{P.}~\bibnamefont{{Astone}}},
  \bibinfo{author}{\bibfnamefont{G.}~\bibnamefont{{Balestri}}},
  \bibinfo{author}{\bibfnamefont{G.}~\bibnamefont{{Ballardin}}},
  \bibnamefont{et~al.}, \bibinfo{journal}{Classical and Quantum Gravity}
  \textbf{\bibinfo{volume}{32}}, \bibinfo{eid}{024001} (\bibinfo{year}{2015}),
  \eprint{1408.3978}.

\bibitem[{\citenamefont{{Akutsu} et~al.}(2021)\citenamefont{{Akutsu}, {Ando},
  {Arai}, {Arai}, {Araki}, {Araya}, {Aritomi}, {Aso}, {Bae}, {Bae}
  et~al.}}]{10.1093/ptep/ptaa125}
\bibinfo{author}{\bibfnamefont{T.}~\bibnamefont{{Akutsu}}},
  \bibinfo{author}{\bibfnamefont{M.}~\bibnamefont{{Ando}}},
  \bibinfo{author}{\bibfnamefont{K.}~\bibnamefont{{Arai}}},
  \bibinfo{author}{\bibfnamefont{Y.}~\bibnamefont{{Arai}}},
  \bibinfo{author}{\bibfnamefont{S.}~\bibnamefont{{Araki}}},
  \bibinfo{author}{\bibfnamefont{A.}~\bibnamefont{{Araya}}},
  \bibinfo{author}{\bibfnamefont{N.}~\bibnamefont{{Aritomi}}},
  \bibinfo{author}{\bibfnamefont{Y.}~\bibnamefont{{Aso}}},
  \bibinfo{author}{\bibfnamefont{S.}~\bibnamefont{{Bae}}},
  \bibinfo{author}{\bibfnamefont{Y.}~\bibnamefont{{Bae}}},
  \bibnamefont{et~al.}, \bibinfo{journal}{Progress of Theoretical and
  Experimental Physics} \textbf{\bibinfo{volume}{2021}}, \bibinfo{eid}{05A101}
  (\bibinfo{year}{2021}), \eprint{2005.05574}.

\bibitem[{\citenamefont{{Abbott}
  et~al.}(2019{\natexlab{b}})\citenamefont{{Abbott}, {Abbott}, {Abbott},
  {Abraham}, {Acernese}, {Ackley}, {Adams}, {Adhikari}, {Adya}, {Affeldt}
  et~al.}}]{Abbott_2019}
\bibinfo{author}{\bibfnamefont{B.~P.} \bibnamefont{{Abbott}}},
  \bibinfo{author}{\bibfnamefont{R.}~\bibnamefont{{Abbott}}},
  \bibinfo{author}{\bibfnamefont{T.~D.} \bibnamefont{{Abbott}}},
  \bibinfo{author}{\bibfnamefont{S.}~\bibnamefont{{Abraham}}},
  \bibinfo{author}{\bibfnamefont{F.}~\bibnamefont{{Acernese}}},
  \bibinfo{author}{\bibfnamefont{K.}~\bibnamefont{{Ackley}}},
  \bibinfo{author}{\bibfnamefont{C.}~\bibnamefont{{Adams}}},
  \bibinfo{author}{\bibfnamefont{R.~X.} \bibnamefont{{Adhikari}}},
  \bibinfo{author}{\bibfnamefont{V.~B.} \bibnamefont{{Adya}}},
  \bibinfo{author}{\bibfnamefont{C.}~\bibnamefont{{Affeldt}}},
  \bibnamefont{et~al.}, \bibinfo{journal}{\apjl}
  \textbf{\bibinfo{volume}{882}}, \bibinfo{eid}{L24}
  (\bibinfo{year}{2019}{\natexlab{b}}), \eprint{1811.12940}.

\bibitem[{\citenamefont{{Abbott}
  et~al.}(2017{\natexlab{a}})\citenamefont{{Abbott}, {Abbott}, {Abbott},
  {Acernese}, {Ackley}, {Adams}, {Adams}, {Addesso}, {Adhikari}, {Adya}
  et~al.}}]{PhysRevLett.119.161101}
\bibinfo{author}{\bibfnamefont{B.~P.} \bibnamefont{{Abbott}}},
  \bibinfo{author}{\bibfnamefont{R.}~\bibnamefont{{Abbott}}},
  \bibinfo{author}{\bibfnamefont{T.~D.} \bibnamefont{{Abbott}}},
  \bibinfo{author}{\bibfnamefont{F.}~\bibnamefont{{Acernese}}},
  \bibinfo{author}{\bibfnamefont{K.}~\bibnamefont{{Ackley}}},
  \bibinfo{author}{\bibfnamefont{C.}~\bibnamefont{{Adams}}},
  \bibinfo{author}{\bibfnamefont{T.}~\bibnamefont{{Adams}}},
  \bibinfo{author}{\bibfnamefont{P.}~\bibnamefont{{Addesso}}},
  \bibinfo{author}{\bibfnamefont{R.~X.} \bibnamefont{{Adhikari}}},
  \bibinfo{author}{\bibfnamefont{V.~B.} \bibnamefont{{Adya}}},
  \bibnamefont{et~al.}, \bibinfo{journal}{\prl} \textbf{\bibinfo{volume}{119}},
  \bibinfo{eid}{161101} (\bibinfo{year}{2017}{\natexlab{a}}),
  \eprint{1710.05832}.

\bibitem[{\citenamefont{{Abbott}
  et~al.}(2017{\natexlab{b}})\citenamefont{{Abbott}, {Abbott}, {Abbott},
  {Acernese}, {Ackley}, {Adams}, {Adams}, {Addesso}, {Adhikari}, {Adya}
  et~al.}}]{k4}
\bibinfo{author}{\bibfnamefont{B.~P.} \bibnamefont{{Abbott}}},
  \bibinfo{author}{\bibfnamefont{R.}~\bibnamefont{{Abbott}}},
  \bibinfo{author}{\bibfnamefont{T.~D.} \bibnamefont{{Abbott}}},
  \bibinfo{author}{\bibfnamefont{F.}~\bibnamefont{{Acernese}}},
  \bibinfo{author}{\bibfnamefont{K.}~\bibnamefont{{Ackley}}},
  \bibinfo{author}{\bibfnamefont{C.}~\bibnamefont{{Adams}}},
  \bibinfo{author}{\bibfnamefont{T.}~\bibnamefont{{Adams}}},
  \bibinfo{author}{\bibfnamefont{P.}~\bibnamefont{{Addesso}}},
  \bibinfo{author}{\bibfnamefont{R.~X.} \bibnamefont{{Adhikari}}},
  \bibinfo{author}{\bibfnamefont{V.~B.} \bibnamefont{{Adya}}},
  \bibnamefont{et~al.}, \bibinfo{journal}{\apjl}
  \textbf{\bibinfo{volume}{848}}, \bibinfo{eid}{L12}
  (\bibinfo{year}{2017}{\natexlab{b}}), \eprint{1710.05833}.

\bibitem[{\citenamefont{{Barack} et~al.}(2019)\citenamefont{{Barack},
  {Cardoso}, {Nissanke}, {Sotiriou}, {Askar}, {Belczynski}, {Bertone}, {Bon},
  {Blas}, {Brito} et~al.}}]{Barack_2019}
\bibinfo{author}{\bibfnamefont{L.}~\bibnamefont{{Barack}}},
  \bibinfo{author}{\bibfnamefont{V.}~\bibnamefont{{Cardoso}}},
  \bibinfo{author}{\bibfnamefont{S.}~\bibnamefont{{Nissanke}}},
  \bibinfo{author}{\bibfnamefont{T.~P.} \bibnamefont{{Sotiriou}}},
  \bibinfo{author}{\bibfnamefont{A.}~\bibnamefont{{Askar}}},
  \bibinfo{author}{\bibfnamefont{C.}~\bibnamefont{{Belczynski}}},
  \bibinfo{author}{\bibfnamefont{G.}~\bibnamefont{{Bertone}}},
  \bibinfo{author}{\bibfnamefont{E.}~\bibnamefont{{Bon}}},
  \bibinfo{author}{\bibfnamefont{D.}~\bibnamefont{{Blas}}},
  \bibinfo{author}{\bibfnamefont{R.}~\bibnamefont{{Brito}}},
  \bibnamefont{et~al.}, \bibinfo{journal}{Classical and Quantum Gravity}
  \textbf{\bibinfo{volume}{36}}, \bibinfo{eid}{143001} (\bibinfo{year}{2019}),
  \eprint{1806.05195}.

\bibitem[{\citenamefont{{Berti} et~al.}(2015)\citenamefont{{Berti}, {Barausse},
  {Cardoso}, {Gualtieri}, {Pani}, {Sperhake}, {Stein}, {Wex}, {Yagi}, {Baker}
  et~al.}}]{Berti_2015}
\bibinfo{author}{\bibfnamefont{E.}~\bibnamefont{{Berti}}},
  \bibinfo{author}{\bibfnamefont{E.}~\bibnamefont{{Barausse}}},
  \bibinfo{author}{\bibfnamefont{V.}~\bibnamefont{{Cardoso}}},
  \bibinfo{author}{\bibfnamefont{L.}~\bibnamefont{{Gualtieri}}},
  \bibinfo{author}{\bibfnamefont{P.}~\bibnamefont{{Pani}}},
  \bibinfo{author}{\bibfnamefont{U.}~\bibnamefont{{Sperhake}}},
  \bibinfo{author}{\bibfnamefont{L.~C.} \bibnamefont{{Stein}}},
  \bibinfo{author}{\bibfnamefont{N.}~\bibnamefont{{Wex}}},
  \bibinfo{author}{\bibfnamefont{K.}~\bibnamefont{{Yagi}}},
  \bibinfo{author}{\bibfnamefont{T.}~\bibnamefont{{Baker}}},
  \bibnamefont{et~al.}, \bibinfo{journal}{Classical and Quantum Gravity}
  \textbf{\bibinfo{volume}{32}}, \bibinfo{eid}{243001} (\bibinfo{year}{2015}),
  \eprint{1501.07274}.

\bibitem[{\citenamefont{{Abbott}
  et~al.}(2017{\natexlab{c}})\citenamefont{{Abbott}, {Abbott}, {Abbott},
  {Acernese}, {Ackley}, {Adams}, {Adams}, {Addesso}, {Adhikari}, {Adya}
  et~al.}}]{Abbott_2017}
\bibinfo{author}{\bibfnamefont{B.~P.} \bibnamefont{{Abbott}}},
  \bibinfo{author}{\bibfnamefont{R.}~\bibnamefont{{Abbott}}},
  \bibinfo{author}{\bibfnamefont{T.~D.} \bibnamefont{{Abbott}}},
  \bibinfo{author}{\bibfnamefont{F.}~\bibnamefont{{Acernese}}},
  \bibinfo{author}{\bibfnamefont{K.}~\bibnamefont{{Ackley}}},
  \bibinfo{author}{\bibfnamefont{C.}~\bibnamefont{{Adams}}},
  \bibinfo{author}{\bibfnamefont{T.}~\bibnamefont{{Adams}}},
  \bibinfo{author}{\bibfnamefont{P.}~\bibnamefont{{Addesso}}},
  \bibinfo{author}{\bibfnamefont{R.~X.} \bibnamefont{{Adhikari}}},
  \bibinfo{author}{\bibfnamefont{V.~B.} \bibnamefont{{Adya}}},
  \bibnamefont{et~al.}, \bibinfo{journal}{\nat} \textbf{\bibinfo{volume}{551}},
  \bibinfo{pages}{85} (\bibinfo{year}{2017}{\natexlab{c}}),
  \eprint{1710.05835}.

\bibitem[{\citenamefont{{Zhu} and {Dong}}(2021)}]{ZhuWei2021ESaT}
\bibinfo{author}{\bibfnamefont{W.}~\bibnamefont{{Zhu}}} \bibnamefont{and}
  \bibinfo{author}{\bibfnamefont{S.}~\bibnamefont{{Dong}}},
  \bibinfo{journal}{\\araa} \textbf{\bibinfo{volume}{59}}, \bibinfo{pages}{291}
  (\bibinfo{year}{2021}), \eprint{2103.02127}.

\bibitem[{\citenamefont{{Annala} et~al.}(2020)\citenamefont{{Annala}, {Gorda},
  {Kurkela}, {N{\"a}ttil{\"a}}, and {Vuorinen}}}]{Annala_2020}
\bibinfo{author}{\bibfnamefont{E.}~\bibnamefont{{Annala}}},
  \bibinfo{author}{\bibfnamefont{T.}~\bibnamefont{{Gorda}}},
  \bibinfo{author}{\bibfnamefont{A.}~\bibnamefont{{Kurkela}}},
  \bibinfo{author}{\bibfnamefont{J.}~\bibnamefont{{N{\"a}ttil{\"a}}}},
  \bibnamefont{and}
  \bibinfo{author}{\bibfnamefont{A.}~\bibnamefont{{Vuorinen}}},
  \bibinfo{journal}{Nature Physics} \textbf{\bibinfo{volume}{16}},
  \bibinfo{pages}{907} (\bibinfo{year}{2020}), \eprint{1903.09121}.

\bibitem[{\citenamefont{{Wierda} et~al.}(2021)\citenamefont{{Wierda}, {Wempe},
  {Hannuksela}, {Koopmans}, and {Van Den Broeck}}}]{k6}
\bibinfo{author}{\bibfnamefont{A.~R. A.~C.} \bibnamefont{{Wierda}}},
  \bibinfo{author}{\bibfnamefont{E.}~\bibnamefont{{Wempe}}},
  \bibinfo{author}{\bibfnamefont{O.~A.} \bibnamefont{{Hannuksela}}},
  \bibinfo{author}{\bibfnamefont{L.~V.~E.} \bibnamefont{{Koopmans}}},
  \bibnamefont{and} \bibinfo{author}{\bibfnamefont{C.}~\bibnamefont{{Van Den
  Broeck}}}, \bibinfo{journal}{\apj} \textbf{\bibinfo{volume}{921}},
  \bibinfo{eid}{154} (\bibinfo{year}{2021}), \eprint{2106.06303}.

\bibitem[{\citenamefont{{Oguri}}(2019)}]{Oguri_2019}
\bibinfo{author}{\bibfnamefont{M.}~\bibnamefont{{Oguri}}},
  \bibinfo{journal}{Reports on Progress in Physics}
  \textbf{\bibinfo{volume}{82}}, \bibinfo{eid}{126901} (\bibinfo{year}{2019}),
  \eprint{1907.06830}.

\bibitem[{\citenamefont{{Liao} et~al.}(2022)\citenamefont{{Liao}, {Biesiada},
  and {Zhu}}}]{Liao:2022gde}
\bibinfo{author}{\bibfnamefont{K.}~\bibnamefont{{Liao}}},
  \bibinfo{author}{\bibfnamefont{M.}~\bibnamefont{{Biesiada}}},
  \bibnamefont{and} \bibinfo{author}{\bibfnamefont{Z.-H.} \bibnamefont{{Zhu}}},
  \bibinfo{journal}{Chinese Physics Letters} \textbf{\bibinfo{volume}{39}},
  \bibinfo{eid}{119801} (\bibinfo{year}{2022}), \eprint{2207.13489}.

\bibitem[{\citenamefont{{Vegetti} and {Koopmans}}(2009)}]{Vegetti_2009}
\bibinfo{author}{\bibfnamefont{S.}~\bibnamefont{{Vegetti}}} \bibnamefont{and}
  \bibinfo{author}{\bibfnamefont{L.~V.~E.} \bibnamefont{{Koopmans}}},
  \bibinfo{journal}{\mnras} \textbf{\bibinfo{volume}{400}},
  \bibinfo{pages}{1583} (\bibinfo{year}{2009}), \eprint{0903.4752}.

\bibitem[{\citenamefont{{Hezaveh} et~al.}(2017)\citenamefont{{Hezaveh},
  {Perreault Levasseur}, and {Marshall}}}]{Hezaveh_2017}
\bibinfo{author}{\bibfnamefont{Y.~D.} \bibnamefont{{Hezaveh}}},
  \bibinfo{author}{\bibfnamefont{L.}~\bibnamefont{{Perreault Levasseur}}},
  \bibnamefont{and} \bibinfo{author}{\bibfnamefont{P.~J.}
  \bibnamefont{{Marshall}}}, \bibinfo{journal}{\nat}
  \textbf{\bibinfo{volume}{548}}, \bibinfo{pages}{555} (\bibinfo{year}{2017}),
  \eprint{1708.08842}.

\bibitem[{\citenamefont{{Nakamura}}(2009)}]{Nakamura_2009}
\bibinfo{author}{\bibfnamefont{K.}~\bibnamefont{{Nakamura}}},
  \bibinfo{journal}{\prd} \textbf{\bibinfo{volume}{80}}, \bibinfo{eid}{124021}
  (\bibinfo{year}{2009}), \eprint{0804.3840}.

\bibitem[{\citenamefont{{Takahashi} and {Nakamura}}(2003)}]{Takahashi_2003}
\bibinfo{author}{\bibfnamefont{R.}~\bibnamefont{{Takahashi}}} \bibnamefont{and}
  \bibinfo{author}{\bibfnamefont{T.}~\bibnamefont{{Nakamura}}},
  \bibinfo{journal}{\apj} \textbf{\bibinfo{volume}{595}}, \bibinfo{pages}{1039}
  (\bibinfo{year}{2003}), \eprint{astro-ph/0305055}.

\bibitem[{\citenamefont{{Dai} et~al.}(2018)\citenamefont{{Dai}, {Li}, {Zackay},
  {Mao}, and {Lu}}}]{Dai_2018}
\bibinfo{author}{\bibfnamefont{L.}~\bibnamefont{{Dai}}},
  \bibinfo{author}{\bibfnamefont{S.-S.} \bibnamefont{{Li}}},
  \bibinfo{author}{\bibfnamefont{B.}~\bibnamefont{{Zackay}}},
  \bibinfo{author}{\bibfnamefont{S.}~\bibnamefont{{Mao}}}, \bibnamefont{and}
  \bibinfo{author}{\bibfnamefont{Y.}~\bibnamefont{{Lu}}},
  \bibinfo{journal}{\prd} \textbf{\bibinfo{volume}{98}}, \bibinfo{eid}{104029}
  (\bibinfo{year}{2018}), \eprint{1810.00003}.

\bibitem[{\citenamefont{{Wu} et~al.}(2019)\citenamefont{{Wu}, {Barnes},
  {Mart{\'\i}nez-Pinedo}, and {Metzger}}}]{PhysRevLett.122.062701}
\bibinfo{author}{\bibfnamefont{M.-R.} \bibnamefont{{Wu}}},
  \bibinfo{author}{\bibfnamefont{J.}~\bibnamefont{{Barnes}}},
  \bibinfo{author}{\bibfnamefont{G.}~\bibnamefont{{Mart{\'\i}nez-Pinedo}}},
  \bibnamefont{and} \bibinfo{author}{\bibfnamefont{B.~D.}
  \bibnamefont{{Metzger}}}, \bibinfo{journal}{\prl}
  \textbf{\bibinfo{volume}{122}}, \bibinfo{eid}{062701} (\bibinfo{year}{2019}),
  \eprint{1808.10459}.

\bibitem[{\citenamefont{{Liao} et~al.}(2017)\citenamefont{{Liao}, {Fan},
  {Ding}, {Biesiada}, and {Zhu}}}]{2017NatCo...8.1148L}
\bibinfo{author}{\bibfnamefont{K.}~\bibnamefont{{Liao}}},
  \bibinfo{author}{\bibfnamefont{X.-L.} \bibnamefont{{Fan}}},
  \bibinfo{author}{\bibfnamefont{X.}~\bibnamefont{{Ding}}},
  \bibinfo{author}{\bibfnamefont{M.}~\bibnamefont{{Biesiada}}},
  \bibnamefont{and} \bibinfo{author}{\bibfnamefont{Z.-H.} \bibnamefont{{Zhu}}},
  \bibinfo{journal}{Nature Communications} \textbf{\bibinfo{volume}{8}},
  \bibinfo{eid}{1148} (\bibinfo{year}{2017}), \eprint{1703.04151}.

\bibitem[{\citenamefont{{Birrer} et~al.}(2024)\citenamefont{{Birrer}, {Millon},
  {Sluse}, {Shajib}, {Courbin}, {Erickson}, {Koopmans}, {Suyu}, and
  {Treu}}}]{k8}
\bibinfo{author}{\bibfnamefont{S.}~\bibnamefont{{Birrer}}},
  \bibinfo{author}{\bibfnamefont{M.}~\bibnamefont{{Millon}}},
  \bibinfo{author}{\bibfnamefont{D.}~\bibnamefont{{Sluse}}},
  \bibinfo{author}{\bibfnamefont{A.~J.} \bibnamefont{{Shajib}}},
  \bibinfo{author}{\bibfnamefont{F.}~\bibnamefont{{Courbin}}},
  \bibinfo{author}{\bibfnamefont{S.}~\bibnamefont{{Erickson}}},
  \bibinfo{author}{\bibfnamefont{L.~V.~E.} \bibnamefont{{Koopmans}}},
  \bibinfo{author}{\bibfnamefont{S.~H.} \bibnamefont{{Suyu}}},
  \bibnamefont{and} \bibinfo{author}{\bibfnamefont{T.}~\bibnamefont{{Treu}}},
  \bibinfo{journal}{\\ssr} \textbf{\bibinfo{volume}{220}}, \bibinfo{eid}{48}
  (\bibinfo{year}{2024}), \eprint{2210.10833}.

\bibitem[{\citenamefont{{Cao} et~al.}(2022)\citenamefont{{Cao}, {Qi}, {Cao},
  {Biesiada}, {Cheng}, and {Zhu}}}]{k9}
\bibinfo{author}{\bibfnamefont{S.}~\bibnamefont{{Cao}}},
  \bibinfo{author}{\bibfnamefont{J.}~\bibnamefont{{Qi}}},
  \bibinfo{author}{\bibfnamefont{Z.}~\bibnamefont{{Cao}}},
  \bibinfo{author}{\bibfnamefont{M.}~\bibnamefont{{Biesiada}}},
  \bibinfo{author}{\bibfnamefont{W.}~\bibnamefont{{Cheng}}}, \bibnamefont{and}
  \bibinfo{author}{\bibfnamefont{Z.-H.} \bibnamefont{{Zhu}}},
  \bibinfo{journal}{\aap} \textbf{\bibinfo{volume}{659}}, \bibinfo{eid}{L5}
  (\bibinfo{year}{2022}), \eprint{2202.08714}.

\bibitem[{\citenamefont{{Owen} and {Sathyaprakash}}(1999)}]{PhysRevD.60.022002}
\bibinfo{author}{\bibfnamefont{B.~J.} \bibnamefont{{Owen}}} \bibnamefont{and}
  \bibinfo{author}{\bibfnamefont{B.~S.} \bibnamefont{{Sathyaprakash}}},
  \bibinfo{journal}{\prd} \textbf{\bibinfo{volume}{60}}, \bibinfo{eid}{022002}
  (\bibinfo{year}{1999}), \eprint{gr-qc/9808076}.

\bibitem[{\citenamefont{{Cornish} and {Littenberg}}(2015)}]{Cornish_2015}
\bibinfo{author}{\bibfnamefont{N.~J.} \bibnamefont{{Cornish}}}
  \bibnamefont{and} \bibinfo{author}{\bibfnamefont{T.~B.}
  \bibnamefont{{Littenberg}}}, \bibinfo{journal}{Classical and Quantum Gravity}
  \textbf{\bibinfo{volume}{32}}, \bibinfo{eid}{135012} (\bibinfo{year}{2015}),
  \eprint{1410.3835}.

\bibitem[{\citenamefont{{Veitch} et~al.}(2015)\citenamefont{{Veitch},
  {Raymond}, {Farr}, {Farr}, {Graff}, {Vitale}, {Aylott}, {Blackburn},
  {Christensen}, {Coughlin} et~al.}}]{PhysRevD.91.042003}
\bibinfo{author}{\bibfnamefont{J.}~\bibnamefont{{Veitch}}},
  \bibinfo{author}{\bibfnamefont{V.}~\bibnamefont{{Raymond}}},
  \bibinfo{author}{\bibfnamefont{B.}~\bibnamefont{{Farr}}},
  \bibinfo{author}{\bibfnamefont{W.}~\bibnamefont{{Farr}}},
  \bibinfo{author}{\bibfnamefont{P.}~\bibnamefont{{Graff}}},
  \bibinfo{author}{\bibfnamefont{S.}~\bibnamefont{{Vitale}}},
  \bibinfo{author}{\bibfnamefont{B.}~\bibnamefont{{Aylott}}},
  \bibinfo{author}{\bibfnamefont{K.}~\bibnamefont{{Blackburn}}},
  \bibinfo{author}{\bibfnamefont{N.}~\bibnamefont{{Christensen}}},
  \bibinfo{author}{\bibfnamefont{M.}~\bibnamefont{{Coughlin}}},
  \bibnamefont{et~al.}, \bibinfo{journal}{\prd} \textbf{\bibinfo{volume}{91}},
  \bibinfo{eid}{042003} (\bibinfo{year}{2015}), \eprint{1409.7215}.

\bibitem[{\citenamefont{{Janquart} et~al.}(2023)\citenamefont{{Janquart},
  {Baka}, {Samajdar}, {Dietrich}, and {Van Den
  Broeck}}}]{Janquart2022AnalysesOO}
\bibinfo{author}{\bibfnamefont{J.}~\bibnamefont{{Janquart}}},
  \bibinfo{author}{\bibfnamefont{T.}~\bibnamefont{{Baka}}},
  \bibinfo{author}{\bibfnamefont{A.}~\bibnamefont{{Samajdar}}},
  \bibinfo{author}{\bibfnamefont{T.}~\bibnamefont{{Dietrich}}},
  \bibnamefont{and} \bibinfo{author}{\bibfnamefont{C.}~\bibnamefont{{Van Den
  Broeck}}}, \bibinfo{journal}{\mnras} \textbf{\bibinfo{volume}{523}},
  \bibinfo{pages}{1699} (\bibinfo{year}{2023}), \eprint{2211.01304}.

\bibitem[{\citenamefont{{Kiendrebeogo}
  et~al.}(2023)\citenamefont{{Kiendrebeogo}, {Farah}, {Foley}, {Gray},
  {Kunert}, {Puecher}, {Toivonen}, {VandenBerg}, {Anand}, {Ahumada}
  et~al.}}]{k40}
\bibinfo{author}{\bibfnamefont{R.~W.} \bibnamefont{{Kiendrebeogo}}},
  \bibinfo{author}{\bibfnamefont{A.~M.} \bibnamefont{{Farah}}},
  \bibinfo{author}{\bibfnamefont{E.~M.} \bibnamefont{{Foley}}},
  \bibinfo{author}{\bibfnamefont{A.}~\bibnamefont{{Gray}}},
  \bibinfo{author}{\bibfnamefont{N.}~\bibnamefont{{Kunert}}},
  \bibinfo{author}{\bibfnamefont{A.}~\bibnamefont{{Puecher}}},
  \bibinfo{author}{\bibfnamefont{A.}~\bibnamefont{{Toivonen}}},
  \bibinfo{author}{\bibfnamefont{R.~O.} \bibnamefont{{VandenBerg}}},
  \bibinfo{author}{\bibfnamefont{S.}~\bibnamefont{{Anand}}},
  \bibinfo{author}{\bibfnamefont{T.}~\bibnamefont{{Ahumada}}},
  \bibnamefont{et~al.}, \bibinfo{journal}{\apj} \textbf{\bibinfo{volume}{958}},
  \bibinfo{eid}{158} (\bibinfo{year}{2023}), \eprint{2306.09234}.

\bibitem[{\citenamefont{{{\c{C}}al{\i}{\c{s}}kan}
  et~al.}(2023)\citenamefont{{{\c{C}}al{\i}{\c{s}}kan}, {Ezquiaga},
  {Hannuksela}, and {Holz}}}]{k41}
\bibinfo{author}{\bibfnamefont{M.}~\bibnamefont{{{\c{C}}al{\i}{\c{s}}kan}}},
  \bibinfo{author}{\bibfnamefont{J.~M.} \bibnamefont{{Ezquiaga}}},
  \bibinfo{author}{\bibfnamefont{O.~A.} \bibnamefont{{Hannuksela}}},
  \bibnamefont{and} \bibinfo{author}{\bibfnamefont{D.~E.}
  \bibnamefont{{Holz}}}, \bibinfo{journal}{\prd}
  \textbf{\bibinfo{volume}{107}}, \bibinfo{eid}{063023} (\bibinfo{year}{2023}),
  \eprint{2201.04619}.

\bibitem[{\citenamefont{{Gabbard} et~al.}(2018)\citenamefont{{Gabbard},
  {Williams}, {Hayes}, and {Messenger}}}]{PhysRevLett.120.141103}
\bibinfo{author}{\bibfnamefont{H.}~\bibnamefont{{Gabbard}}},
  \bibinfo{author}{\bibfnamefont{M.}~\bibnamefont{{Williams}}},
  \bibinfo{author}{\bibfnamefont{F.}~\bibnamefont{{Hayes}}}, \bibnamefont{and}
  \bibinfo{author}{\bibfnamefont{C.}~\bibnamefont{{Messenger}}},
  \bibinfo{journal}{\prl} \textbf{\bibinfo{volume}{120}}, \bibinfo{eid}{141103}
  (\bibinfo{year}{2018}), \eprint{1712.06041}.

\bibitem[{\citenamefont{{Green} and {Gair}}(2020)}]{Green_2021}
\bibinfo{author}{\bibfnamefont{S.~R.} \bibnamefont{{Green}}} \bibnamefont{and}
  \bibinfo{author}{\bibfnamefont{J.}~\bibnamefont{{Gair}}},
  \bibinfo{journal}{arXiv e-prints} \bibinfo{eid}{arXiv:2008.03312}
  (\bibinfo{year}{2020}), \eprint{2008.03312}.

\bibitem[{\citenamefont{{Chua} and
  {Vallisneri}}(2020)}]{PhysRevLett.124.041102}
\bibinfo{author}{\bibfnamefont{A.~J.~K.} \bibnamefont{{Chua}}}
  \bibnamefont{and}
  \bibinfo{author}{\bibfnamefont{M.}~\bibnamefont{{Vallisneri}}},
  \bibinfo{journal}{\prl} \textbf{\bibinfo{volume}{124}}, \bibinfo{eid}{041102}
  (\bibinfo{year}{2020}), \eprint{1909.05966}.

\bibitem[{\citenamefont{{Gao} et~al.}(2023)\citenamefont{{Gao}, {Liao}, {Yang},
  and {Zhu}}}]{Gao2023}
\bibinfo{author}{\bibfnamefont{Z.}~\bibnamefont{{Gao}}},
  \bibinfo{author}{\bibfnamefont{K.}~\bibnamefont{{Liao}}},
  \bibinfo{author}{\bibfnamefont{L.}~\bibnamefont{{Yang}}}, \bibnamefont{and}
  \bibinfo{author}{\bibfnamefont{Z.-H.} \bibnamefont{{Zhu}}},
  \bibinfo{journal}{\mnras} \textbf{\bibinfo{volume}{526}},
  \bibinfo{pages}{682} (\bibinfo{year}{2023}), \eprint{2304.13967}.

\bibitem[{\citenamefont{{Collett}}(2015)}]{Collett2015}
\bibinfo{author}{\bibfnamefont{T.~E.} \bibnamefont{{Collett}}},
  \bibinfo{journal}{\apj} \textbf{\bibinfo{volume}{811}}, \bibinfo{eid}{20}
  (\bibinfo{year}{2015}), \eprint{1507.02657}.

\bibitem[{\citenamefont{{Magare} et~al.}(2024)\citenamefont{{Magare}, {More},
  and {Choudhary}}}]{Magare:2024wje}
\bibinfo{author}{\bibfnamefont{S.}~\bibnamefont{{Magare}}},
  \bibinfo{author}{\bibfnamefont{A.}~\bibnamefont{{More}}}, \bibnamefont{and}
  \bibinfo{author}{\bibfnamefont{S.}~\bibnamefont{{Choudhary}}},
  \bibinfo{journal}{\mnras} \textbf{\bibinfo{volume}{535}},
  \bibinfo{pages}{990} (\bibinfo{year}{2024}), \eprint{2403.02994}.

\bibitem[{\citenamefont{{Goyal} et~al.}(2021)\citenamefont{{Goyal},
  {Harikrishnan}, {Kapadia}, and {Ajith}}}]{Goyal2021}
\bibinfo{author}{\bibfnamefont{S.}~\bibnamefont{{Goyal}}},
  \bibinfo{author}{\bibfnamefont{D.}~\bibnamefont{{Harikrishnan}}},
  \bibinfo{author}{\bibfnamefont{S.~J.} \bibnamefont{{Kapadia}}},
  \bibnamefont{and} \bibinfo{author}{\bibfnamefont{P.}~\bibnamefont{{Ajith}}},
  \bibinfo{journal}{\prd} \textbf{\bibinfo{volume}{104}}, \bibinfo{eid}{124057}
  (\bibinfo{year}{2021}), \eprint{2106.12466}.

\bibitem[{\citenamefont{{Kim} et~al.}(2020)\citenamefont{{Kim}, {Lee}, {Yuen},
  {Akseli Hannuksela}, and {Li}}}]{Kim:2020xkm}
\bibinfo{author}{\bibfnamefont{K.}~\bibnamefont{{Kim}}},
  \bibinfo{author}{\bibfnamefont{J.}~\bibnamefont{{Lee}}},
  \bibinfo{author}{\bibfnamefont{R.~S.~H.} \bibnamefont{{Yuen}}},
  \bibinfo{author}{\bibfnamefont{O.}~\bibnamefont{{Akseli Hannuksela}}},
  \bibnamefont{and} \bibinfo{author}{\bibfnamefont{T.~G.~F.}
  \bibnamefont{{Li}}}, \bibinfo{journal}{arXiv e-prints}
  \bibinfo{eid}{arXiv:2010.12093} (\bibinfo{year}{2020}), \eprint{2010.12093}.

\bibitem[{\citenamefont{{Lin} et~al.}(2023)\citenamefont{{Lin}, {Zhang}, {Dai},
  {Huang}, and {Mei}}}]{Lin2023}
\bibinfo{author}{\bibfnamefont{X.-y.} \bibnamefont{{Lin}}},
  \bibinfo{author}{\bibfnamefont{J.-d.} \bibnamefont{{Zhang}}},
  \bibinfo{author}{\bibfnamefont{L.}~\bibnamefont{{Dai}}},
  \bibinfo{author}{\bibfnamefont{S.-J.} \bibnamefont{{Huang}}},
  \bibnamefont{and} \bibinfo{author}{\bibfnamefont{J.}~\bibnamefont{{Mei}}},
  \bibinfo{journal}{\prd} \textbf{\bibinfo{volume}{108}}, \bibinfo{eid}{064020}
  (\bibinfo{year}{2023}), \eprint{2304.04800}.

\bibitem[{\citenamefont{{Pagano} et~al.}(2020)\citenamefont{{Pagano},
  {Hannuksela}, and {Li}}}]{Pagano2020}
\bibinfo{author}{\bibfnamefont{G.}~\bibnamefont{{Pagano}}},
  \bibinfo{author}{\bibfnamefont{O.~A.} \bibnamefont{{Hannuksela}}},
  \bibnamefont{and} \bibinfo{author}{\bibfnamefont{T.~G.~F.}
  \bibnamefont{{Li}}}, \bibinfo{journal}{\aap} \textbf{\bibinfo{volume}{643}},
  \bibinfo{eid}{A167} (\bibinfo{year}{2020}), \eprint{2006.12879}.

\bibitem[{\citenamefont{{Sheng} et~al.}(2022)\citenamefont{{Sheng}, {C},
  {Choi}, {Sharpnack}, and {Jones}}}]{Sheng2022}
\bibinfo{author}{\bibfnamefont{S.}~\bibnamefont{{Sheng}}},
  \bibinfo{author}{\bibfnamefont{K.~V.~G.} \bibnamefont{{C}}},
  \bibinfo{author}{\bibfnamefont{C.~P.} \bibnamefont{{Choi}}},
  \bibinfo{author}{\bibfnamefont{J.}~\bibnamefont{{Sharpnack}}},
  \bibnamefont{and} \bibinfo{author}{\bibfnamefont{T.}~\bibnamefont{{Jones}}},
  \bibinfo{journal}{arXiv e-prints} \bibinfo{eid}{arXiv:2210.11681}
  (\bibinfo{year}{2022}), \eprint{2210.11681}.

\bibitem[{\citenamefont{{Keerthi Vasan} et~al.}(2023)\citenamefont{{Keerthi
  Vasan}, {Sheng}, {Jones}, {Choi}, and {Sharpnack}}}]{KeerthiVasan2023}
\bibinfo{author}{\bibfnamefont{G.~C.} \bibnamefont{{Keerthi Vasan}}},
  \bibinfo{author}{\bibfnamefont{S.}~\bibnamefont{{Sheng}}},
  \bibinfo{author}{\bibfnamefont{T.}~\bibnamefont{{Jones}}},
  \bibinfo{author}{\bibfnamefont{C.~P.} \bibnamefont{{Choi}}},
  \bibnamefont{and}
  \bibinfo{author}{\bibfnamefont{J.}~\bibnamefont{{Sharpnack}}},
  \bibinfo{journal}{\mnras} \textbf{\bibinfo{volume}{524}},
  \bibinfo{pages}{5368} (\bibinfo{year}{2023}), \eprint{2211.00047}.

\bibitem[{\citenamefont{{Chakraborty} and
  {Mukherjee}}(2024)}]{2024MNRAS.532.4842C}
\bibinfo{author}{\bibfnamefont{A.}~\bibnamefont{{Chakraborty}}}
  \bibnamefont{and}
  \bibinfo{author}{\bibfnamefont{S.}~\bibnamefont{{Mukherjee}}},
  \bibinfo{journal}{\mnras} \textbf{\bibinfo{volume}{532}},
  \bibinfo{pages}{4842} (\bibinfo{year}{2024}), \eprint{2403.03982}.

\bibitem[{\citenamefont{{Cheung} et~al.}(2021)\citenamefont{{Cheung}, {Gais},
  {Hannuksela}, and {Li}}}]{10.1093/mnras/stab579}
\bibinfo{author}{\bibfnamefont{M.~H.~Y.} \bibnamefont{{Cheung}}},
  \bibinfo{author}{\bibfnamefont{J.}~\bibnamefont{{Gais}}},
  \bibinfo{author}{\bibfnamefont{O.~A.} \bibnamefont{{Hannuksela}}},
  \bibnamefont{and} \bibinfo{author}{\bibfnamefont{T.~G.~F.}
  \bibnamefont{{Li}}}, \bibinfo{journal}{\mnras}
  \textbf{\bibinfo{volume}{503}}, \bibinfo{pages}{3326} (\bibinfo{year}{2021}),
  \eprint{2012.07800}.

\bibitem[{\citenamefont{Finder et~al.}(2024)\citenamefont{Finder, Amoyal,
  Treister, and Freifeld}}]{finder2024wavelet}
\bibinfo{author}{\bibfnamefont{S.~E.} \bibnamefont{Finder}},
  \bibinfo{author}{\bibfnamefont{R.}~\bibnamefont{Amoyal}},
  \bibinfo{author}{\bibfnamefont{E.}~\bibnamefont{Treister}}, \bibnamefont{and}
  \bibinfo{author}{\bibfnamefont{O.}~\bibnamefont{Freifeld}}, in
  \emph{\bibinfo{booktitle}{European Conference on Computer Vision}}
  (\bibinfo{year}{2024}), \eprint{2407.05848}.

\bibitem[{\citenamefont{{He} et~al.}(2016)\citenamefont{{He}, {Zhang}, {Ren},
  and {Sun}}}]{he2016deep}
\bibinfo{author}{\bibfnamefont{K.}~\bibnamefont{{He}}},
  \bibinfo{author}{\bibfnamefont{X.}~\bibnamefont{{Zhang}}},
  \bibinfo{author}{\bibfnamefont{S.}~\bibnamefont{{Ren}}}, \bibnamefont{and}
  \bibinfo{author}{\bibfnamefont{J.}~\bibnamefont{{Sun}}}, in
  \emph{\bibinfo{booktitle}{2016 IEEE Conference on Computer Vision and Pattern
  Recognition (CVPR}} (\bibinfo{year}{2016}), p.~\bibinfo{pages}{1},
  \eprint{1512.03385}.

\bibitem[{\citenamefont{{Grespan} and {Biesiada}}(2023)}]{k14}
\bibinfo{author}{\bibfnamefont{M.}~\bibnamefont{{Grespan}}} \bibnamefont{and}
  \bibinfo{author}{\bibfnamefont{M.}~\bibnamefont{{Biesiada}}},
  \bibinfo{journal}{Universe} \textbf{\bibinfo{volume}{9}}, \bibinfo{eid}{200}
  (\bibinfo{year}{2023}).

\bibitem[{\citenamefont{{Meena} and {Bagla}}(2020)}]{k43}
\bibinfo{author}{\bibfnamefont{A.~K.} \bibnamefont{{Meena}}} \bibnamefont{and}
  \bibinfo{author}{\bibfnamefont{J.~S.} \bibnamefont{{Bagla}}},
  \bibinfo{journal}{\mnras} \textbf{\bibinfo{volume}{492}},
  \bibinfo{pages}{1127} (\bibinfo{year}{2020}), \eprint{1903.11809}.

\bibitem[{\citenamefont{{Sun} et~al.}(2023)\citenamefont{{Sun}, {Shi}, {Zhang},
  and {Mei}}}]{Sun_2023}
\bibinfo{author}{\bibfnamefont{S.}~\bibnamefont{{Sun}}},
  \bibinfo{author}{\bibfnamefont{C.}~\bibnamefont{{Shi}}},
  \bibinfo{author}{\bibfnamefont{J.-d.} \bibnamefont{{Zhang}}},
  \bibnamefont{and} \bibinfo{author}{\bibfnamefont{J.}~\bibnamefont{{Mei}}},
  \bibinfo{journal}{\prd} \textbf{\bibinfo{volume}{107}}, \bibinfo{eid}{044023}
  (\bibinfo{year}{2023}), \eprint{2207.13009}.

\bibitem[{\citenamefont{{Urrutia} and {Vaskonen}}(2022)}]{Urrutia2021}
\bibinfo{author}{\bibfnamefont{J.}~\bibnamefont{{Urrutia}}} \bibnamefont{and}
  \bibinfo{author}{\bibfnamefont{V.}~\bibnamefont{{Vaskonen}}},
  \bibinfo{journal}{\mnras} \textbf{\bibinfo{volume}{509}},
  \bibinfo{pages}{1358} (\bibinfo{year}{2022}), \eprint{2109.03213}.

\bibitem[{\citenamefont{{Mukherjee} et~al.}(2021)\citenamefont{{Mukherjee},
  {Broadhurst}, {Diego}, {Silk}, and {Smoot}}}]{Mukherjee2021}
\bibinfo{author}{\bibfnamefont{S.}~\bibnamefont{{Mukherjee}}},
  \bibinfo{author}{\bibfnamefont{T.}~\bibnamefont{{Broadhurst}}},
  \bibinfo{author}{\bibfnamefont{J.~M.} \bibnamefont{{Diego}}},
  \bibinfo{author}{\bibfnamefont{J.}~\bibnamefont{{Silk}}}, \bibnamefont{and}
  \bibinfo{author}{\bibfnamefont{G.~F.} \bibnamefont{{Smoot}}},
  \bibinfo{journal}{\mnras} \textbf{\bibinfo{volume}{506}},
  \bibinfo{pages}{3751} (\bibinfo{year}{2021}), \eprint{2106.00392}.

\bibitem[{\citenamefont{{Madau} and {Fragos}}(2017)}]{Madau2016}
\bibinfo{author}{\bibfnamefont{P.}~\bibnamefont{{Madau}}} \bibnamefont{and}
  \bibinfo{author}{\bibfnamefont{T.}~\bibnamefont{{Fragos}}},
  \bibinfo{journal}{\apj} \textbf{\bibinfo{volume}{840}}, \bibinfo{eid}{39}
  (\bibinfo{year}{2017}), \eprint{1606.07887}.

\bibitem[{\citenamefont{{Liao} et~al.}(2020)\citenamefont{{Liao}, {Tian}, and
  {Ding}}}]{Liao2020}
\bibinfo{author}{\bibfnamefont{K.}~\bibnamefont{{Liao}}},
  \bibinfo{author}{\bibfnamefont{S.}~\bibnamefont{{Tian}}}, \bibnamefont{and}
  \bibinfo{author}{\bibfnamefont{X.}~\bibnamefont{{Ding}}},
  \bibinfo{journal}{\mnras} \textbf{\bibinfo{volume}{495}},
  \bibinfo{pages}{2002} (\bibinfo{year}{2020}), \eprint{2001.07891}.

\bibitem[{\citenamefont{{Basak} et~al.}(2022)\citenamefont{{Basak}, {Ganguly},
  {Haris}, {Kapadia}, {Mehta}, and {Ajith}}}]{Basak2021}
\bibinfo{author}{\bibfnamefont{S.}~\bibnamefont{{Basak}}},
  \bibinfo{author}{\bibfnamefont{A.}~\bibnamefont{{Ganguly}}},
  \bibinfo{author}{\bibfnamefont{K.}~\bibnamefont{{Haris}}},
  \bibinfo{author}{\bibfnamefont{S.}~\bibnamefont{{Kapadia}}},
  \bibinfo{author}{\bibfnamefont{A.~K.} \bibnamefont{{Mehta}}},
  \bibnamefont{and} \bibinfo{author}{\bibfnamefont{P.}~\bibnamefont{{Ajith}}},
  \bibinfo{journal}{\apjl} \textbf{\bibinfo{volume}{926}}, \bibinfo{eid}{L28}
  (\bibinfo{year}{2022}), \eprint{2109.06456}.

\bibitem[{\citenamefont{{Zhou} et~al.}(2023)\citenamefont{{Zhou}, {Li}, {Liao},
  and {Huang}}}]{Zhou2022}
\bibinfo{author}{\bibfnamefont{H.}~\bibnamefont{{Zhou}}},
  \bibinfo{author}{\bibfnamefont{Z.}~\bibnamefont{{Li}}},
  \bibinfo{author}{\bibfnamefont{K.}~\bibnamefont{{Liao}}}, \bibnamefont{and}
  \bibinfo{author}{\bibfnamefont{Z.}~\bibnamefont{{Huang}}},
  \bibinfo{journal}{\mnras} \textbf{\bibinfo{volume}{518}},
  \bibinfo{pages}{149} (\bibinfo{year}{2023}), \eprint{2206.13128}.

\bibitem[{\citenamefont{{Shan} et~al.}(2025)\citenamefont{{Shan}, {Hu}, {Chen},
  and {Cai}}}]{Shan:2023ngi}
\bibinfo{author}{\bibfnamefont{X.}~\bibnamefont{{Shan}}},
  \bibinfo{author}{\bibfnamefont{B.}~\bibnamefont{{Hu}}},
  \bibinfo{author}{\bibfnamefont{X.}~\bibnamefont{{Chen}}}, \bibnamefont{and}
  \bibinfo{author}{\bibfnamefont{R.-G.} \bibnamefont{{Cai}}},
  \bibinfo{journal}{Nature Astronomy} \textbf{\bibinfo{volume}{9}},
  \bibinfo{pages}{916} (\bibinfo{year}{2025}), \eprint{2301.06117}.

\bibitem[{\citenamefont{{Lin} et~al.}(2025)\citenamefont{{Lin}, {Wang}, {Zhou},
  {Li}, {Liao}, and {Zhu}}}]{Lin:2025mpx}
\bibinfo{author}{\bibfnamefont{X.-Y.} \bibnamefont{{Lin}}},
  \bibinfo{author}{\bibfnamefont{X.-J.} \bibnamefont{{Wang}}},
  \bibinfo{author}{\bibfnamefont{H.}~\bibnamefont{{Zhou}}},
  \bibinfo{author}{\bibfnamefont{Z.}~\bibnamefont{{Li}}},
  \bibinfo{author}{\bibfnamefont{K.}~\bibnamefont{{Liao}}}, \bibnamefont{and}
  \bibinfo{author}{\bibfnamefont{Z.-H.} \bibnamefont{{Zhu}}},
  \bibinfo{journal}{arXiv e-prints} \bibinfo{eid}{arXiv:2508.13577}
  (\bibinfo{year}{2025}), \eprint{2508.13577}.

\bibitem[{\citenamefont{{Suyamprakasam}
  et~al.}(2025)\citenamefont{{Suyamprakasam}, {Harikumar}, {Cieciel{\'a}g},
  {Figura}, {Bejger}, and {Biesiada}}}]{Suyamprakasam:2025pvk}
\bibinfo{author}{\bibfnamefont{S.}~\bibnamefont{{Suyamprakasam}}},
  \bibinfo{author}{\bibfnamefont{S.}~\bibnamefont{{Harikumar}}},
  \bibinfo{author}{\bibfnamefont{P.}~\bibnamefont{{Cieciel{\'a}g}}},
  \bibinfo{author}{\bibfnamefont{P.}~\bibnamefont{{Figura}}},
  \bibinfo{author}{\bibfnamefont{M.}~\bibnamefont{{Bejger}}}, \bibnamefont{and}
  \bibinfo{author}{\bibfnamefont{M.}~\bibnamefont{{Biesiada}}},
  \bibinfo{journal}{arXiv e-prints} \bibinfo{eid}{arXiv:2503.21845}
  (\bibinfo{year}{2025}), \eprint{2503.21845}.

\bibitem[{\citenamefont{{Liao} et~al.}(2019)\citenamefont{{Liao}, {Biesiada},
  and {Fan}}}]{Liao:2019aqq}
\bibinfo{author}{\bibfnamefont{K.}~\bibnamefont{{Liao}}},
  \bibinfo{author}{\bibfnamefont{M.}~\bibnamefont{{Biesiada}}},
  \bibnamefont{and} \bibinfo{author}{\bibfnamefont{X.-L.} \bibnamefont{{Fan}}},
  \bibinfo{journal}{\apj} \textbf{\bibinfo{volume}{875}}, \bibinfo{eid}{139}
  (\bibinfo{year}{2019}), \eprint{1903.06612}.

\bibitem[{\citenamefont{LeCun et~al.}(2015)\citenamefont{LeCun, Bengio, and
  Hinton}}]{lecun2015deep}
\bibinfo{author}{\bibfnamefont{Y.}~\bibnamefont{LeCun}},
  \bibinfo{author}{\bibfnamefont{Y.}~\bibnamefont{Bengio}}, \bibnamefont{and}
  \bibinfo{author}{\bibfnamefont{G.}~\bibnamefont{Hinton}},
  \bibinfo{journal}{Nature} \textbf{\bibinfo{volume}{521}},
  \bibinfo{pages}{436} (\bibinfo{year}{2015}).

\bibitem[{\citenamefont{Krizhevsky et~al.}(2012)\citenamefont{Krizhevsky,
  Sutskever, and Hinton}}]{krizhevsky2012imagenet}
\bibinfo{author}{\bibfnamefont{A.}~\bibnamefont{Krizhevsky}},
  \bibinfo{author}{\bibfnamefont{I.}~\bibnamefont{Sutskever}},
  \bibnamefont{and} \bibinfo{author}{\bibfnamefont{G.~E.}
  \bibnamefont{Hinton}}, in \emph{\bibinfo{booktitle}{Advances in Neural
  Information Processing Systems}} (\bibinfo{year}{2012}), pp.
  \bibinfo{pages}{1097--1105}.

\bibitem[{\citenamefont{Dong et~al.}(2015)\citenamefont{Dong, Loy, He, and
  Tang}}]{dong2015image}
\bibinfo{author}{\bibfnamefont{C.}~\bibnamefont{Dong}},
  \bibinfo{author}{\bibfnamefont{C.~C.} \bibnamefont{Loy}},
  \bibinfo{author}{\bibfnamefont{K.}~\bibnamefont{He}}, \bibnamefont{and}
  \bibinfo{author}{\bibfnamefont{X.}~\bibnamefont{Tang}}, in
  \emph{\bibinfo{booktitle}{Proceedings of the IEEE Conference on Computer
  Vision and Pattern Recognition}} (\bibinfo{year}{2015}), pp.
  \bibinfo{pages}{166--174}.

\bibitem[{\citenamefont{Szegedy et~al.}(2015)\citenamefont{Szegedy, Liu, Jia,
  Sermanet, Reed, Anguelov, Erhan, Vanhoucke, and
  Rabinovich}}]{szegedy2015going}
\bibinfo{author}{\bibfnamefont{C.}~\bibnamefont{Szegedy}},
  \bibinfo{author}{\bibfnamefont{W.}~\bibnamefont{Liu}},
  \bibinfo{author}{\bibfnamefont{Y.}~\bibnamefont{Jia}},
  \bibinfo{author}{\bibfnamefont{P.}~\bibnamefont{Sermanet}},
  \bibinfo{author}{\bibfnamefont{S.}~\bibnamefont{Reed}},
  \bibinfo{author}{\bibfnamefont{D.}~\bibnamefont{Anguelov}},
  \bibinfo{author}{\bibfnamefont{D.}~\bibnamefont{Erhan}},
  \bibinfo{author}{\bibfnamefont{V.}~\bibnamefont{Vanhoucke}},
  \bibnamefont{and}
  \bibinfo{author}{\bibfnamefont{A.}~\bibnamefont{Rabinovich}}, in
  \emph{\bibinfo{booktitle}{Proceedings of the IEEE Conference on Computer
  Vision and Pattern Recognition}} (\bibinfo{year}{2015}), pp.
  \bibinfo{pages}{1--9}, \eprint{1409.4842}.

\bibitem[{\citenamefont{Dosovitskiy et~al.}(2021)\citenamefont{Dosovitskiy,
  Beyer, Kolesnikov, Weissenborn, Zhai, Unterthiner, Dehghani, Minderer,
  Heigold, Gelly et~al.}}]{vit2021}
\bibinfo{author}{\bibfnamefont{A.}~\bibnamefont{Dosovitskiy}},
  \bibinfo{author}{\bibfnamefont{L.}~\bibnamefont{Beyer}},
  \bibinfo{author}{\bibfnamefont{A.}~\bibnamefont{Kolesnikov}},
  \bibinfo{author}{\bibfnamefont{D.}~\bibnamefont{Weissenborn}},
  \bibinfo{author}{\bibfnamefont{X.}~\bibnamefont{Zhai}},
  \bibinfo{author}{\bibfnamefont{T.}~\bibnamefont{Unterthiner}},
  \bibinfo{author}{\bibfnamefont{M.}~\bibnamefont{Dehghani}},
  \bibinfo{author}{\bibfnamefont{M.}~\bibnamefont{Minderer}},
  \bibinfo{author}{\bibfnamefont{G.}~\bibnamefont{Heigold}},
  \bibinfo{author}{\bibfnamefont{S.}~\bibnamefont{Gelly}},
  \bibnamefont{et~al.}, in \emph{\bibinfo{booktitle}{International Conference
  on Learning Representations}} (\bibinfo{year}{2021}), \eprint{2010.11929}.

\bibitem[{\citenamefont{Nitz et~al.}(2024)\citenamefont{Nitz, Harry, Brown,
  Biwer, Willis, Canton, Capano, Dent, Pekowsky, Davies et~al.}}]{alex}
\bibinfo{author}{\bibfnamefont{A.}~\bibnamefont{Nitz}},
  \bibinfo{author}{\bibfnamefont{I.}~\bibnamefont{Harry}},
  \bibinfo{author}{\bibfnamefont{D.}~\bibnamefont{Brown}},
  \bibinfo{author}{\bibfnamefont{C.~M.} \bibnamefont{Biwer}},
  \bibinfo{author}{\bibfnamefont{J.}~\bibnamefont{Willis}},
  \bibinfo{author}{\bibfnamefont{T.~D.} \bibnamefont{Canton}},
  \bibinfo{author}{\bibfnamefont{C.}~\bibnamefont{Capano}},
  \bibinfo{author}{\bibfnamefont{T.}~\bibnamefont{Dent}},
  \bibinfo{author}{\bibfnamefont{L.}~\bibnamefont{Pekowsky}},
  \bibinfo{author}{\bibfnamefont{G.~S.~C.} \bibnamefont{Davies}},
  \bibnamefont{et~al.}, \emph{\bibinfo{title}{gwastro/pycbc: v2.3.3 release of
  pycbc}} (\bibinfo{year}{2024}).

\bibitem[{\citenamefont{Kingma and Ba}(2014)}]{Kingma2014AdamAM}
\bibinfo{author}{\bibfnamefont{D.~P.} \bibnamefont{Kingma}} \bibnamefont{and}
  \bibinfo{author}{\bibfnamefont{J.}~\bibnamefont{Ba}}, \bibinfo{journal}{CoRR}
  \textbf{\bibinfo{volume}{abs/1412.6980}} (\bibinfo{year}{2014}).

\bibitem[{\citenamefont{Loshchilov and Hutter}(2017)}]{sgdr2017}
\bibinfo{author}{\bibfnamefont{I.}~\bibnamefont{Loshchilov}} \bibnamefont{and}
  \bibinfo{author}{\bibfnamefont{F.}~\bibnamefont{Hutter}}, in
  \emph{\bibinfo{booktitle}{Proc. of ICLR}} (\bibinfo{year}{2017}),
  \eprint{1608.03983}.

\bibitem[{\citenamefont{{Fawcett}}(2006)}]{k33}
\bibinfo{author}{\bibfnamefont{T.}~\bibnamefont{{Fawcett}}},
  \bibinfo{journal}{Pattern Recognition Letters} \textbf{\bibinfo{volume}{27}},
  \bibinfo{pages}{861} (\bibinfo{year}{2006}).

\end{thebibliography}
\end{document}